\documentclass[a4paper,12pt,oneside]{report} 

\usepackage{EMU_Thesis}
\usepackage[a4paper,left=4cm,right=2.5cm,top=2.5cm,bottom=2.5cm,footskip=.8cm]{geometry}

\pgfplotsset{width=10cm,compat=1.9}
\usepgfplotslibrary{external}

\title{Interacting Null Sources in Different Geometries}
\author{Chia-Li Hsieh}
\degree{Doctor of Philosophy}
\inof{in}
\program{Physics}
\defenseyear{2024}
\defensemonth{June}
\Dept{Physics}
\InstituteDirector{Prof. Dr. Ali Hakan Ulusoy}
\DeptChair{Prof. Dr. İzzet Sakallı}
\supervisor{Prof. Dr. Mustafa Halilsoy}
\examiner{Prof. Dr.}{Mustafa}{Halilsoy}
\examiner{Prof. Dr.}{Muzaffer}{Adak}
\examiner{Prof. Dr.}{Özay}{Gürtuğ}
\examiner{Prof. Dr.}{İzzet}{Sakallı}
\examiner{Prof. Dr.}{Seyedhabibollah}{Mazharimousavi}

\begin{document}

\pagenumbering{roman}

\tolerance=1
\emergencystretch=\maxdimen
\hyphenpenalty=10000
\hbadness=10000

\makethesistitle 
\makeapprovalpage

\begin{abstract}
Firstly, we introduce fundamental mathematical techniques, followed by an exploration of three distinct topics: the Callan-Giddings-Harvey-Strominger (CGHS) model in 1+1-dimensional spacetime, the formation of astrophysical jets in Schwarzschild-like black holes, and collisions and confinement phenomena in third-order Lovelock gravity. Here are the key features.

In the CGHS model, we investigate the collision of ghost fields within the dilaton background geometry, observing the formation and dissolution of wormholes by inserting and removing the ghost fields, respectively. We say that this process resembles a cosmological-scale analogue of Feynman diagrams.

Next, we explore the emergence of non-zero expectation values of bumblebee fields due to Lorentz symmetry breaking. This alteration in the energy-momentum tensor necessitates the inclusion of a potential vacuum, resulting in a shift of the vacuum solution towards Schwarzchild-like black holes with a scaling factor $l$. This scaling factor facilitates discussions on the collision of null sources, leading to the formation of impulsive null shells and satisfying the type-D condition in the interacting region. When $l$ approaches zero, jet-like formations vanish, transforming the problem into one involving colliding gravitational waves, which is isometric to the Schwarzschild geometry. Moreover, our method can be applied to any metric resembling Schwarzschild-like configurations. We aim to enhance our model by incorporating additional physical factors such as extra polarizations or electromagnetic fields.

Finally, our examination extends to the realm of 4-dimensional third-order Lovelock gravity. We observe that each particle possessing finite energy experiences confinement within the metric time interval extending from negative to positive infinity. Moreover, this finding does not accommodate flat rotation curves. Additionally, when collisions occur within the background of this metric, intriguingly, we observe impulsive Weyl curvatures along the null boundaries subsequent to the collision.

\end{abstract}

\begin{ozet}
Öncelikle konu ile ilgili temel matematik tekniklerini veriyoruz, ardından Callan-Giddings-Harvey-Strominger (CGHS) 1+1 boyutlu uzay-zaman modeli, astrofiziksel jetlerin oluşumu, Schwarzschild benzeri kara delikler, çarpışmalar ve hapsedilme olayları ile üçüncü dereceden Lovelock yerçekimi olmak üzere üç farklı konuyu ele alıyoruz.

CGHS modelinde Feynman diyagramlarının kozmolojik ölçekteki bir benzerine benzettiğimiz dilaton içindeki hayalet alanların çarpışmasını araştırarak  arka plan geometrisini, solucan deliklerinin oluşumunu ve çözünmesini sırasıyla hayalet alanları ekleme ve kaldırma ile gözlemliyoruz. 

Daha sonra, Bumblebee çekim modeli alanlarının sıfırdan farklı beklenti değerlerinin ortaya çıkışını araştırıyoruz. Lorentz simetrisinin kırılması nedeniyle, enerji-momentum tensöründeki bu değişiklik potansiyel bir vakumun dahil edilmesini gerektirir, bu da vakumun kaymasına neden olur ve ölçeklendirme faktörü l olan Schwarzchild benzeri kara delik çözümü verir. Bu ölçeklendirme faktörü, boş kaynakların çarpışmasıyla ilgili tartışmaları kolaylaştırır ve yeni oluşumlara yol açar. Dürtüsel boş kabukların ve etkileşimli bölgedeki D tipi koşulun karşılanması, l sıfıra yaklaştığında jet benzeri oluşumlar ortadan kaybolmasına ve sorun Schwarzschild'e eşdeğer olan, çarpışan yerçekimi dalgalarını içeren bir dalga uzayı oluşmasına neden olur. Ayrıca yöntemimiz, Schwarzschild benzeri farklı konfigürasyonlara  modelimizi geliştirerek uygulamayı amaçlıyor, öğrneğin; ekstra polarizasyonlar veya elektromanyetik alanlar gibi ek fiziksel faktörler.

Son olarak incelememiz 4 boyutlu üçüncü dereceden Lovelock yerçekimi alanına kadar uzanıyor. Her parçacığın sonlu enerji deneyimine sahip olduğunu gözlemliyoruz. Burada negatiften pozitife uzanan sonsuz metrik zaman aralığı içindeki sınırlama kalıcıdır. Üstelik bu bulgu düz dönüş eğrilerini kapsamamaktadır. Ek olarak, bu metriğin arka planında çarpışmalar meydana geldiğinde, ilginç bir şekilde, ışıksal sınırları boyunca dürtüsel Weyl eğriliklerini gözlemliyoruz.

\noindent \textbf{Anahtar Kelimeler}: 1+1 Boyut, CGHS modeli, Bumblebee yerçekimi, Astrofiziksel jet,
Lovelock yerçekimi, Işıksal zar (kabuk) çarpışması.

\end{ozet}

\begin{notitlededication} 
\vspace{\fill} 
\begin{center}
{\calligra... Dedicated to my parents} 

\end{center}
\vspace{\fill}
\end{notitlededication}

\begin{acknowledgements} 

I would like to express my gratitude towards Prof. MUSTAFA HALİLSOY who helped me in completing my study and contributed to preparing my thesis and also to the faculty of physics department, especially for chairman Prof. İZZET SAKALLI who helps me in everything in detailed. Additionally, I would like to thank Prof. Muzaffer Adak, Prof. Özay GÜRTUĞ and Prof. Seyedhabibollah Mazharimousavi for attending my thesis defense and the help from my close friedns, Mert and Erdem. When life gives you lemons, make lemonade. I know that to pursuit the PhD is not easy, but I don't know that it is so complicated for not just study but also troubles in my life. I want to thank everyone who encourages me not to give up in my tough time. And my dad and mom, you are backing nothing but everything. 


\end{acknowledgements}


\tableofcontents

\listoftables

\listoffigures





\DTLnewdb{symbols}
\addsymbol{$c$}{Speed of Light in a Vacuum Inertial Frame}
\addsymbol{$G$}{Gravitational Constant}
\addsymbol[b]{$k$}{The Strength of the Monopole}
\addsymbol[b]{$l$}{Lorentz Symmetry Breaking Constant}
\addsymbol[b]{$\kappa$}{$8\pi G$}

\DTLsort*{Acronym}{symbols}

\DTLnewdb{abbreviation}
\addabbreviation{NP}{Newman-Penrose}
\addabbreviation{CGHS}{Callan-Giddings-Harvey-Strominger }
\addabbreviation{TOLG}{Third-order Lovelock Gravity}
\addabbreviation{GB}{Gauss-Bonnet}
\addabbreviation{CF}{Conformally Flat}
\addabbreviation{FRC}{ Flat Rotation Curve}

\DTLsort*{Acronym}{abbreviation}


\begin{symbolsandabbreviations}
\end{symbolsandabbreviations}




\chapter{INTRODUCTION}\label{ch:introduction}
\pagenumbering{arabic}
In this thesis, we explore gravitational dynamics and astrophysical phenomena by examining 3 separate cases, including the dilatonic gravity in 1+1-dimension, the astrophysical jets by the metric coming from bumblebee gravity and the geometry from the pure third-order Lovelock gravity. We mainly rely on the colliding technique and Newman-Penrose (NP) formalism to deal with these 3 topics.

Here are the features we have found.
1)In the dilatonic gravity, we develop the phenomena arising from the Callan-Giddings-Harvey-Strominger (CGHS) model, a two-dimensional toy model of gravity. By examining the interplay between colliding ghost fields, wormhole formation, and black hole dynamics in the lineland geometry, we uncover fascinating insights into the nature of spacetime.
2) We examine the implications of the bumblebee gravity model or equivalently the global monopole model for the possible dynamics of astrophysical jets emanating from black holes.
3)The third-order pure Lovelock theory leads to particle confinement, and the collision of null sources to the emergence of static spacetime structures.

In this chapter we should acquaint with the mathematical tools of collision by step functions first. Then we shall explain how we introduce the ideas of CGHS model, bumblebee gravity and Lovelock gravity theory in this chapter.
Furthermore, some mathematical preliminaries would be introduced which will help us go through the whole article smoothly. We shall understand the collision pictures in the null coordinates. This diagrams provide the understanding of the structure of spacetime before and after the collision. For the collisions here are between ghost-ghost fields, real-real fields, and null-dust waves. Among them, We exploit the step functions to create the colliding processes.

Then we shall solidify the mathematical tools of NP formalism by using differential forms, in contrary to using tensors in Griffiths' book \cite{griffiths2016colliding}. The approach of differential forms makes the calculation more economical. Because some unwanted components with 0 can be avoided. And the most important, Ricci tensor and Weyl tensor are indeed derived form NP formalism. These tensors are of great importance when we are going to discuss the gravitational fields.

Therefore in this chapter,there includes the introduction of CGHS model, Bumblebee gravity and Lovelock gravity, then our thesis is organized as follows. In chapter II, we discuss the collision of ghost fields in 1+1-D dilatonic background geometry. We discuss colliding null matter with a specific stress tensor in chapter III. In chapter IV, the confinement and null-source collision in a particular 4-dimensional third-order Lovelock gravity is imposed. We complete the thesis with discussions and conclusions in chapter V.

\section{Mathematical Preliminaries: Step Functions in Spacetime Structure}\label{sec:generalinstructions}
A step function, $\theta(x)$, also known as the Heaviside step function, is defined as:

\begin{equation}
\theta(x) =
\left\{
    \begin{array}{lr}
        1, \qquad   \text{if } &  x \geq 0\\
        0, \qquad   \text{if } &  x < 0
    \end{array}
    \right\}
\end{equation}
In other words, it is a function that is 0 for all negative values of its argument and 1 for all non-negative values. The step function is normally used in signal processing, to model sudden changes or transitions, like switching on a system or activating a process at a particular time. Here, we say that the interaction takes place after $ x\geq 0$, and nothing happened when
$x < 0$. Besides, one may regard $\theta(x)$ as shock waves in the metric and its derivative $\delta(x)$ the impulsive waves. We shall consider an example for showing the whole idea much well. In the metric \cite{griffiths2016colliding}

\begin{equation}
  ds^2=2dudv-(1-u\theta(u))^2dx^2-(1+u\theta(u))^2dy^2,
\end{equation}
where $du=\frac{1}{\sqrt{2}}(dt+dz)$ and $dv=\frac{1}{\sqrt{2}}(dt-dz)$ are called null coordinates, specialized in describing the path of massless particles, like electromagnetic and gravitational waves. $u=\frac{1}{\sqrt{2}}(t+z)$ responds for incoming waves and $v=\frac{1}{\sqrt{2}}(t-z)$ for outgoing. So we can turn on the metric for $u \geq0$, or spacetime remains flat when $u < 0$. The gravitational waves in the $u\geq0$ area is $\Psi_4=\delta(u)$, a impulsive wave (we shall explain the method of how to derive $\Psi_s$ in the next section). To describe the incoming wave along $v$, we should swap $u$ for $v$. In the $v\geq0$ region, the gravitational wave would be $\Psi_0=\delta(v)$. So one can imagine that there would be regions with empty only, $\Psi_4=\delta(u)$ only and $\Psi_0=\delta(v)$ only. And there should be a special region that we turn on both of $\theta(u)$ and $\theta(v)$. We say that the waves collide at $u=0$, $v=0$ and develop into a interacting region and a more complicated metric. In this way, we can construct the interacting picture in Figure 1.1.

\begin{figure}
  \centering
  \includegraphics[width=0.5\textwidth]{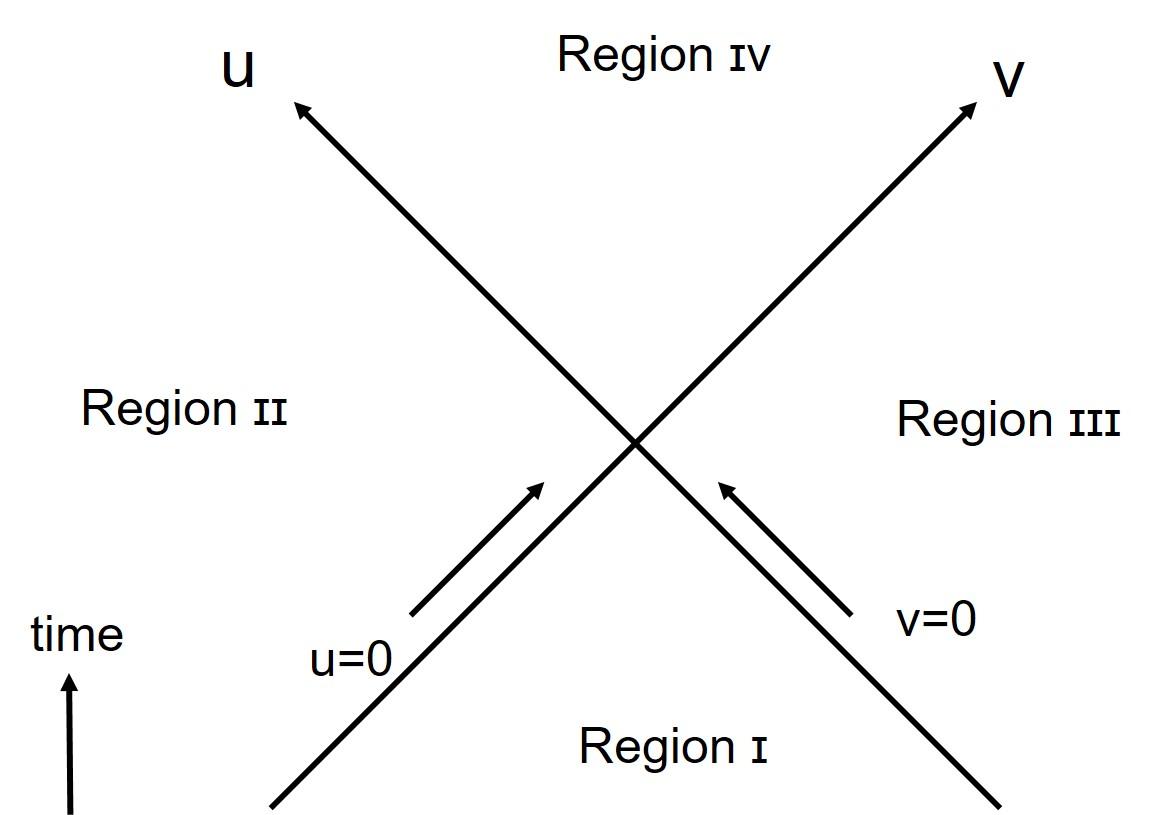}
  \caption{\centering Two waves collide at $u=0$ and $v=0$, since we switch on the interaction at this point by step functions.}\label{f2}
\end{figure}

One can see that in region I, $u<0$ and $v<0$, there is nothing but flat metric

\begin{equation}
  ds^2=2dudv-dx^2-dy^2.
\end{equation}

In region II, $u\geq0$ and $v<0$, we have

\begin{equation}
  ds^2=2dudv-(1-u)^2dx^2-(1+u)^2dy^2,
\end{equation}

and
\begin{equation}
  ds^2=2dudv-(1-v)^2dx^2-(1+v)^2dy^2,
\end{equation}
in region III. But in region IV, $u>0$ and $v>0$ , the line element is not easy. Please see Khan and Penrose \cite{khan1971scattering}. We are not going to solve it but state the result,

\begin{equation}
 \begin{split}
  ds^2=
  -(1-u^2-v^2)(\frac{1-u\sqrt{1-v^2}-v\sqrt{1-u^2}}{1+u\sqrt{1-v^2}+v\sqrt{1-u^2}}dx^2+\frac{1+u\sqrt{1-v^2}+v\sqrt{1-u^2}}{1-u\sqrt{1-v^2}-v\sqrt{1-u^2}} dy^2)\\
  +2\frac{(1-u^2-v^2)^{3/2}}{\sqrt{1-u^2}\sqrt{1-v^2}(uv+\sqrt{1-u^2}\sqrt{1-v^2})^2}dudv\\.
  \end{split}
\end{equation}

Also, the gravitational fields would be

\begin{equation}
 \begin{split}
 \Psi_0=\frac{1}{\sqrt{1-u^2}}\delta(v)+\frac{3u\sqrt{1-u^2}(uv+\sqrt{1-u^2}\sqrt{1-v^2})}{(1-v^2)(1-u^2-v^2)^2} \\
 \Psi_4=\frac{1}{\sqrt{1-v^2}}\delta(u)+\frac{3v\sqrt{1-v^2}(uv+\sqrt{1-u^2}\sqrt{1-v^2})}{(1-u^2)(1-u^2-v^2)^2}\\
 \Psi_2=\frac{(uv+\sqrt{1-u^2}\sqrt{1-v^2})^2}{\sqrt{1-u^2}\sqrt{1-v^2}(1-u^2-v^2)^2}-\frac{uv}{(1-u^2-v^2)^2} \\
  \end{split}
\end{equation}

The $\Psi_0$ and $\Psi_4$ vary a little and develop tails. In addition, the feature of interacting wave, $\Psi_2$, is also developed. One can see that there are coordinate singularities, $u=1$ and $v=1$ in region II and III, which are removable \cite{matzner1984metaphysics}. However, in the region IV, there exists a curvature singularity, $u^2+v^2=1$. Therefore, we may plot the spacetime structure of Khan-Penrose metric as follows. If someone travels in the path A, he will hit the curvature singularity and ends there. But if he travels in the path B, he may pass the coordinate singularity and prevent the curvature singularity. Throughout the thesis, we shall stick to this technique.

\begin{figure}
  \centering
  \includegraphics[width=1\textwidth]{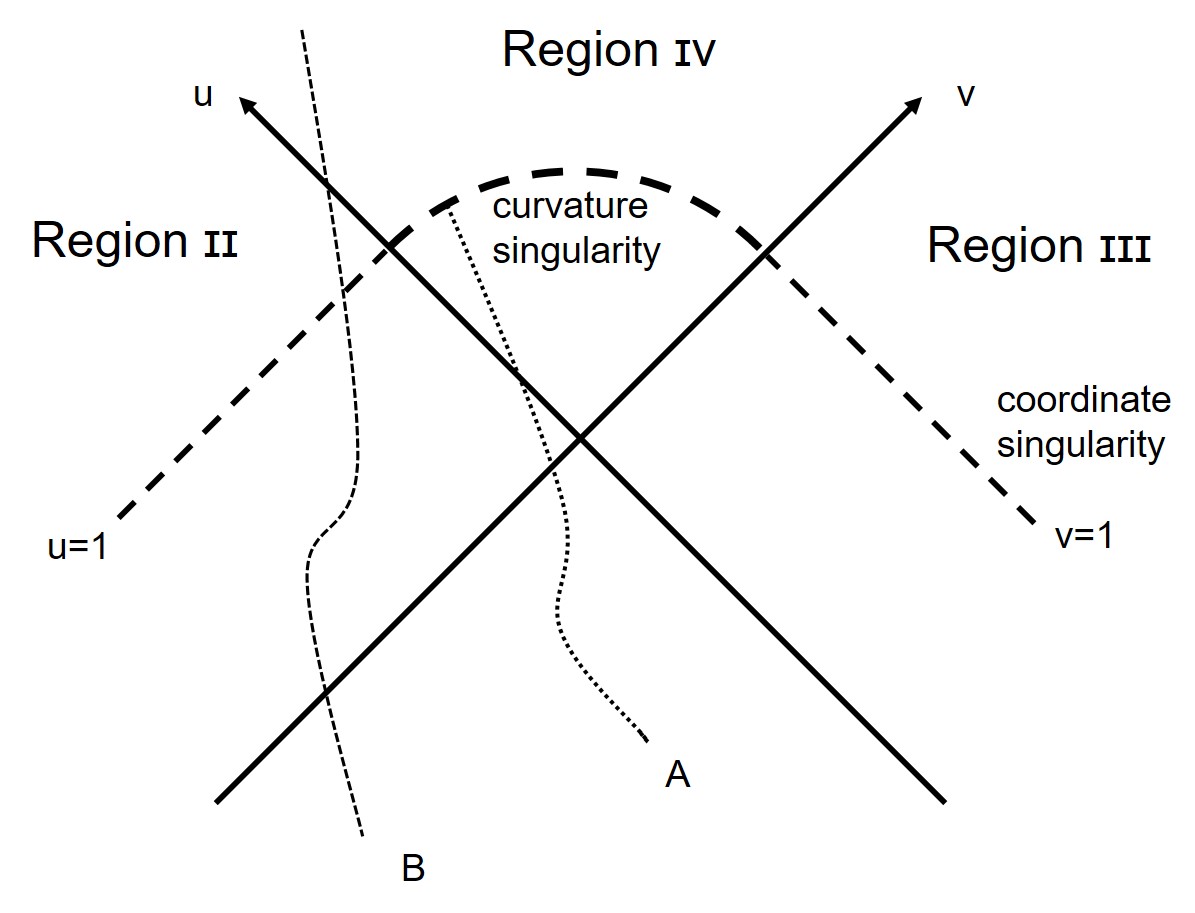}
  \caption{\centering There are curvature singularity and coordinate singularity in the Khan-Penrose spacetime.}\label{f2}
\end{figure}

\section{The 1+1-dimensional Dilaton Gravity}
\subsection{The CGHS Model }

The 1+1-dimensional dilaton gravity is a simplified model used in theoretical physics, particularly in the context of string theory and black hole physics. It serves as a useful toy model for exploring certain aspects of gravity and quantum field theory in curved spacetime.

The gravitational dynamics are described by a two-dimensional spacetime with a metric tensor $g_{\mu\nu}$ and a scalar field $\phi$ called the dilaton field. The line element and action for this theory typically take the form:

\begin{equation}
    ds^2=-2e^{2\phi}dudv,
\end{equation}
and
\begin{equation}
  S=\int d^2x\sqrt{-g}[e^{-2\phi}(R+4(\nabla\phi)^2+4\lambda^2],
\end{equation}
where
R is the Ricci scalar, representing the curvature of spacetime, $(\nabla\phi)^2$ represents the kinetic energy of the dilaton field, and $\lambda$ the cosmological constant. This is the Callan-Giddings-Harvey-Strominger (CGHS) model of which the theory are governed by the equations of motion obtained by varying the action with respect to the metric tensor and the dilaton field. 
We can call this 1+1-D the "lineland" that provides some calculation simplicities that 2+1-D and 3+1-D models can not reach.
\subsection{The Features} 
Unlike the flat Minkowski vacuum, the background is supported by the cosmological constant $\lambda$, that is, even in the vacuum spacetime the background is not flat but bended by $\lambda$. Therefore, in the cosmological-constant vacuum of higher dimensions, higher than 2, the collision waves is a challenging problem, the incoming waves become distorted due to the nonlinear coupling. Furthermore, we put real fields and ghost fields into the Lagrangian. Then as previous section mentioned, we choose $u$, $v$ as null coordinate and make collision at $u=0=v$ by setting step functions. The trick is to choose proper functions for the extra fields. Since the functions involve second derivatives, $sin(\theta(u))$ and $sinh(\theta(u))$ are not good choices, they shall give unwanted $\delta(u)$ after the derivatives. We have to use $cos(\theta(u))$ or $cosh(\theta(u))$ as our colliding waves. Interestingly, we have found that the collision of ghost-ghost fields creates wormholes. This view point of particle model coincide the wormhole geometry of Einstein and Rosen \cite{einstein1935particle} and similar to the quantum colliding case of $\gamma$ photon giving birth of particle and anti-particle pairs. Our model is that wormholes will collapse to a black hole whenever the perturbation is removed, just like the Feynman diagram that perturbations creates virtual particles but restored after unperturbed condition.

\section{Colliding Waves in Bumblebee Gravity }
Further, we would like to discuss the dynamical region of waves in static metric. In Bumblebee gravity, there is a static solution with a Lorentz symmetry breaking constant $l$. (We shall discuss the Bumblebee gravity of Lorentz symmetry breaking in great detail in chapter 3). The waves are derived in this metric and therefore we can make the waves colliding. The metric reads

\begin{equation}
  ds^2=(1-\frac{2m}{r})dt^2-(1+l)\frac{dr^2}{(1-\frac{2m}{r})}-r^2(d\theta^2+\sin^2\theta d\phi^2),
\end{equation}

where $m$= mass, $l$= Lorentz symmetry breaking factor. This metric can be transformed into the dynamical form

\begin{equation}
  ds^2=(1+\tau)^2(2dudv-cos^2(\frac{u-v}{\sqrt{1-2k}})dy^2)-(\frac{1-\tau}{1+\tau})dx^2,
\end{equation}

where $l=\frac{2k}{1-2k}$= constant and $\tau=sin(u+v)$. Again, we shall discuss this transformation in chapter 3. One can see that we have set the null coordinates $(u, v)$ here, which means that we can let $u\rightarrow u(\theta)$ and $v\rightarrow v(\theta)$ as we have shown in section 1.1. Such that the waves would collide when the step functions $\theta(u)$ and $\theta(v)$ are turned on.

Learning this technique, we try to describe the "astrophysical jet", a jet erupted from a accretion disk around black holes, by this Schwarzchild-like black hole solution. But one should know that we only collide null sources, which is spinless and chargeless. The collision between spin and charged waves would be challenged. This is a corner stone for understanding the astrophysical phenomena like jet.  

\section{Confinement and Null-source Collision of the 
Third-order Lovelock Gravity}

We are interested in the pure third-order Lovelock gravity, that is, we strip off some unwanted parameters of other orders. We shall derive the line element

\begin{equation}
ds^{2}=-\left( 1\pm 2ar\right) dt^{2}+\frac{dr^{2}}{\left( 1\pm 2ar\right) }%
+r^{2}d\Omega ^{2}  
\end{equation}%

This metric is a particular form of the Mannheim-Kazaras spacetime. We find the radial geodesics confinement under this geometry and the flat
rotation curve condition is not satisfied.

we transform this static metric into the double
null-coordinates ($u,v$),

\begin{equation}
ds^{2}=\frac{1}{F^{2}}\left[ -2dudv+a^{2}\cos ^{2}\left( u+v\right)
dx^{2}+\cos ^{2}\left( u-v\right) dy^{2}\right].
\end{equation}%

Again, we discuss the collision of this metric. Notably, there exist impulsive Weyl curvatures along the null boundaries after the collision.

\section{Penrose Diagram}
\subsection{Penrose Diagram of Minkowski Metric}

Penrose diagrams are tools for visualizing and understanding the sturcture of spacetime. It's a two-dimensional diagram that maps the universe to a finite area. Please see Hawking and Ellis \cite{hawking2023large}.
To illustrate the diagram, we may consider the Minkowski flat spacetime,

\begin{equation}
  ds^2=-dt^2+dr^2+r^2(d\theta^2+sin^2\theta d\phi^2)=-dudv+\frac{1}{4}(u-v)^2(d\theta^2+sin^2\theta d\phi^2),
\end{equation}

where the null coordinate, $u=t-r$, $v=t+r$. For mapping infinity to be finite, we set $u=tan (q)$, $v=tan (p)$, such that

\begin{equation}
  ds^2=\frac{1}{cos^2p cos^2q}(-dpdq+\frac{1}{4}sin^2(p-q)d\Omega^2),
\end{equation}

where $-\frac{\pi}{2}<p\leq q<\frac{\pi}{2}$. We can always find a replacement that $T=p+q$, $R=p-q$, and $0\leq R < \pi$, $|T| < \pi-R$, such that

\begin{equation}
  ds^2=\frac{1}{4cos^2p cos^2q}(-dT^2+dR^2+sin^2Rd\Omega^2),
\end{equation}

Let $\theta=constant$, $\phi=constant\;\rightarrow d\Omega^2=0$. Since $R=tan^{-1}(t+r)-tan^{-1}(t-r)$, $T=tan^{-1}(t+r)+tan^{-1}(t-r)$, one can find that the universe is now mapped into R-T coordinate with

\begin{equation}
  \begin{array}{lr}
    1) t=constant\;, r\to\infty,\;R=\pi,\;T=0,\; denoted\;i^0,\;$for spatial infinity$  \\
    2) r=constant\;, t\to\infty,\;R=0,\;T=\pi,\; denoted\;i^+,\;$for future timelike infinity$ ) \\
    3) t=constant\;, r\to-\infty,\;R=0, \; T=-\pi,\; denoted\;i^-,\;$for past timelike infinity$ \\
    4) I^+,\;I^-, \; $correspond to r-t=constant$, r+t=\infty $\;and r+t=constant$,\;r-t=-\infty.

  \end{array}
\end{equation}

$I^+$ and $I^-$ are the future and pass null infinity, respectively. These are null surfaces. The outgoing radial null geodesics end at $I^+$. While, the incoming radial null geodesics started at$I^-$. We make a mapping in Figure 1.3.

\begin{figure}
  \centering
  \includegraphics[width=1\textwidth]{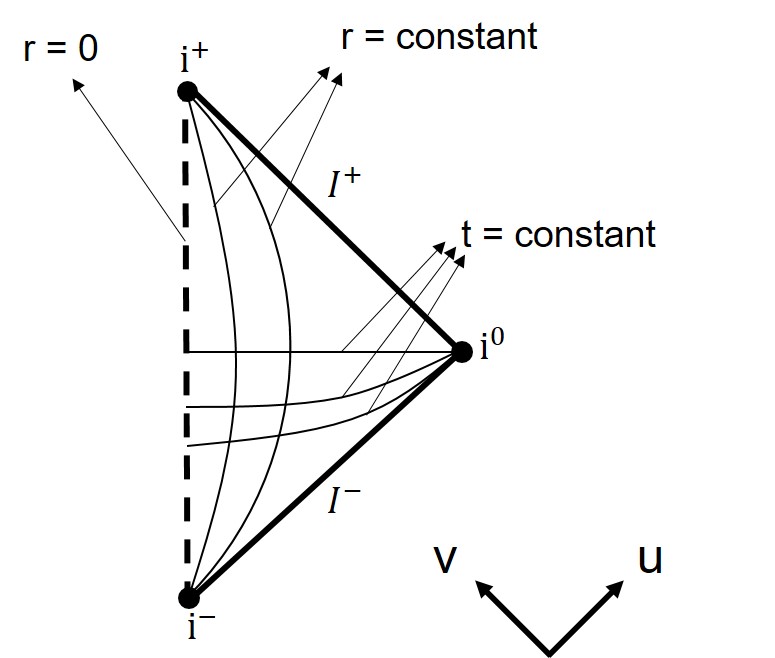}
  \caption{\centering The Penrose diagram of Minkowski spacetime.}\label{f2}
\end{figure}

\subsection{Penrose Diagram of Schwarzschild Metric}
After the standard case, we should move on to another standard case, which will appear again and again in the thesis but with varied forms, that is, the Schwarzschild metric  $ds^2=-(1-\frac{2m}{r})dt^2+(1-\frac{2m}{r})^{-1}dr^2+r^2d\Omega^2$. One can transform the Schwarzschild metric into the Kruskal-Szekeres metric, see \cite{hawking2023large, szekeres1960singularities},

\begin{equation}
  ds^2=-\frac{16m^3}{r}e^{\frac{-r}{2m}}(dudv)+r^2d\Omega^2.
\end{equation}

We choose $\theta=constant$, $\phi=constant$ and let $U=\tan^{-1}(\frac{u}{\sqrt{2m}})$ and $V=\tan^{-1}(\frac{v}{\sqrt{2m}})$, the range would be $-\pi/2 < U <+\pi/2, -\pi/2 < V <+\pi/2, -\pi < U+V <+\pi $. So that we can draw the Penrose diagram as follows in figure 1.4.

\begin{figure}
  \centering
  \includegraphics[width=1\textwidth]{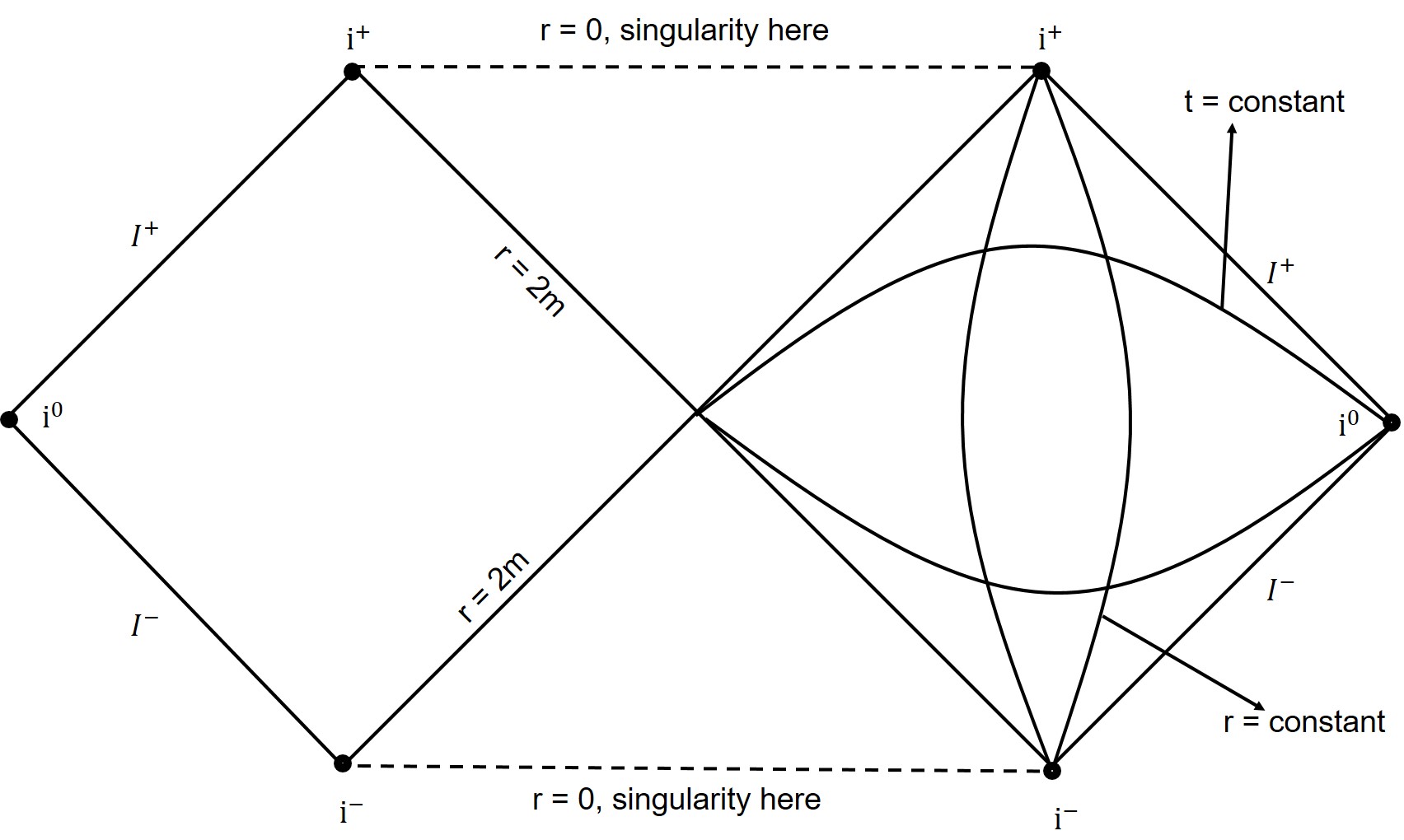}
  \caption{\centering The Penrose diagram of Schwarzschild spacetime.}\label{f2}
\end{figure}

\section{Mathematical Preliminaries: Exterior Product, Derivative, and Forms}

\subsection{Exterior Algebra}
The exterior product, also known as the wedge product, is used to define exterior differential forms.
We should briefly introduce the exterior product and the derivative for further used in Newman-Penrose formalism.

Calculus is working on tangent space with basis $\frac{\partial}{\partial x^i}$. We can always find a dual space with basis $dx^j$, called 1-forms, which will map the vector into real numbers, with $<dx^j, \frac{\partial}{\partial x^i}>=\delta^j_i$. The basis 1-forms are asymmetric under wedge product, that is, $A\wedge B=-B\wedge A$. The scalar function f would be called 0-form.  Then the exterior derivative, d, is defined as $df=\frac{\partial f}{\partial x^i}dx^i$. The exterior derivative of a function is called a 1-form. Therefore, we can label it as $w^i=f(x^i)dx^i$: 1-form. $dw^i=\frac{\partial f}{\partial x^j}dx^j\wedge dx^i$: 2-form. The exterior derivative of 1-form is 2-form, and so on and so forth. Furthermore, d of d is always zero, that is, $ddw=0$. As a result, if $dw=0$, which implies $w=d\eta$ in a simply connected space (see \cite{hawking2023large, thorne2000gravitation}). This is pretty much like the curl of the gradient of a function is always 0. $\nabla\times\bar{V}=0\; \longrightarrow \bar{V}=\nabla \phi$. We say a form is exact if it is of the form $w=d\eta$, and a form is closed if $d\Omega=0$. One may clearly see all the relations in the table 1.1.

\begin{table}[!htbp]
\caption{ \centering Exterior Algebra}
\label{table:nonlin}
\begin{tabular}{|l|}
\hline
d: exterior derivative  \\ \hline\hline
$A\wedge B=-B\wedge A$   \\ \hline
$w^i=f(x^i)dx^i$: 1-form    \\ \hline
$dw^i=\frac{\partial f}{\partial x^j}dx^j\wedge dx^i$: 2-form   \\ \hline
$\omega=d\eta$: exact form  \\ \hline
$d\Omega=0$: closed form \\ \hline

\end{tabular}
\end{table}

Since the wedge product can be applied to any dimensions, we should figure out some further properties of the form space, $\Lambda$.

\begin{equation}
  \begin{array}{lr}
    1) d:\Lambda^k\to\Lambda^{k+1}  \\
    2) d(\omega+\eta)=d\omega+d\eta\;   (\omega, \eta\:  $are the same dimension form$) \\
    3) d(\omega\wedge\eta)=d\omega\wedge\eta+(-1)^n\omega\wedge d\eta\;  (\omega:n-form, \eta:k-form)\\
  \end{array}
\end{equation}

\subsection{Differential Forms}
Then, we may consider the potential 1-form $A=A_\mu dx^\mu=\phi dt-A_x dx-A_y dy-A_z dz$. Let $F=dA$, we have
\begin{equation*}
  \begin{array}{lr}
  F=dA=\partial_\nu A_\mu dx^\nu\wedge dx^\mu\\
  =-\partial_\nu A_\mu dx^\mu\wedge dx^\nu=\frac{1}{2}dA+\frac{1}{2}dA\\
  =\frac{1}{2}(\partial_\mu A_\nu-\partial_\nu A_\mu)dx^\nu\wedge dx^\mu\\
  =\frac{1}{2}F_{\mu\nu}dx^\mu\wedge dx^\nu.
  \end{array}
\end{equation*}

Also, we can set $\theta$ as a 0-form. Such that

\begin{equation}
  A'=A+d\theta\;\rightarrow dA'=d(A+d\theta)=dA,
\end{equation}

which is gauge invariance.

We know that $E=-\nabla\phi-\frac{\partial \overrightarrow{\rm A}}{\partial t}$ and $B=\nabla \times \overrightarrow{\rm A}$. Therefore,

\begin{equation*}
  \begin{array}{lr}
  F=(E_xdx+E_ydy+E_zdz)\wedge dt+B_xdy\wedge dz+B_ydz\wedge dx+B_zdx\wedge dy.\\

  \rightarrow dF=\frac{\partial E_x}{\partial y}dy\wedge dx\wedge dt+\frac{\partial E_x}{\partial z}dz\wedge dx\wedge dt+...\\
  =(\frac{\partial E_y}{\partial x}-\frac{\partial E_x}{\partial y}+\frac{\partial B_z}{\partial t})dx\wedge dy\wedge dt+(\nabla\cdot \overrightarrow{\rm B})dx\wedge dy\wedge dz+...

  \end{array}
\end{equation*}

$\therefore dF=0$ leads to

\begin{equation}
  \begin{array}{lr}
  \nabla \times \overrightarrow{\rm E}+\frac{\partial \overrightarrow{\rm B}}{\partial t}=0\\\
  \nabla\cdot\overrightarrow{\rm B}=0,

  \end{array}
\end{equation}

the homogeneous Maxwell equations. Interestingly,

\begin{equation}
  \begin{array}{lr}
  dF=\frac{1}{2}\partial_\lambda F_{\mu\nu} dx^\lambda\wedge dx^\mu\wedge dx^\nu\\
  =\frac{1}{2}\partial_\mu F_{\nu\lambda} dx^\mu\wedge dx^\nu\wedge dx^\lambda\\
  =\frac{1}{3}dF+\frac{1}{3}dF+\frac{1}{3}dF\\
  =\frac{1}{6}(\partial_\lambda F_{\mu\nu}+\partial_\nu F_{\lambda\mu}+\partial_\mu F_{\nu\lambda}) dx^\lambda\wedge dx^\mu\wedge dx^\nu\\
  =0\\
  \rightarrow \partial_\lambda F_{\mu\nu}+\partial_\nu F_{\lambda\mu}+\partial_\mu F_{\nu\lambda}=0,
  \end{array}
\end{equation}

the Bianchi identity is derived.
However, we are just in the halfway. To get the other pair of Maxwell equations, one should consider the dual forms, which are of great importance to find the dual of $F_{\mu\nu}$, especially in some complicated geometries.

In Minkowski spacetime of $(+,-,-,-)$ signature, the dual forms of basis 2-forms are defined as

\begin{equation}
  \begin{array}{lr}
  *(dx\wedge dt)=-(dy\wedge dz),\;*(dy\wedge dt)=-(dz\wedge dx),\;*(dz\wedge dt)=-(dx\wedge dy),\;\\
  *(dy\wedge dz)=(dx\wedge dt),\;*(dz\wedge dx)=(dy\wedge dt),\;*(dx\wedge dy)=(dz\wedge dt),\;\\
  *(dx\wedge dy\wedge dz)=dt,\;*(dt\wedge dy\wedge dz)=dz,\;*(dt\wedge dz\wedge dx)=dy,\;*(dt\wedge dx\wedge dy)=dz,\;
  \end{array}
\end{equation}

where the * is the Hodge operator to define the dual forms, see \cite{thorne2000gravitation}. Note that different signature may pick up minus signs. We shall learn how to find the $\pm$ signs latter. In this way, we can write down the dual of $F$.

\begin{equation*}
  \begin{array}{lr}
  *F=-E_xdy\wedge dz-E_ydz\wedge dz-E_zdx\wedge dy+B_xdx\wedge dt+B_ydy\wedge dt+B_zdz\wedge dt.\\
  $And define the current 3-forms$\\
  J=(j_xdy\wedge dz+j_ydz\wedge dx+j_zdx\wedge dy)\wedge dt-\rho dx\wedge dy\wedge dz.\\

  \end{array}
\end{equation*}

One will find that $d*F=J$ by matching all the terms. This is the inhomogenous Maxwell equations, $\nabla\times\overrightarrow{\rm B}-\frac{\partial\overrightarrow{\rm E}}{\partial t}=\overrightarrow{\rm j}\; and\; \nabla\cdot\overrightarrow{\rm E}=\rho$. Interestingly,   $d(d*F)=0, \rightarrow dJ=0,\rightarrow \frac{\partial\rho}{\partial t}+\nabla\cdot\overrightarrow{\rm J}=0$, we have the continuity equation.

\subsection{Dual Forms}
Since we have seen the importance of dual forms, we should learn how to derive them. Latter on, we shall face the tetrad in Newman-Penrose. Without dual forms, it would be complicated to get the dual of tetrad. In previous section, we can notice that $\omega \in \Lambda^k,\rightarrow *\omega\in\Lambda^{n-k},(n=4)$, which is that the total dimension is n=4, if the forms are k dimensions, the dual forms have (n-k) dimensions. Therefore, we can see, for example, $*(dx\wedge dy\wedge dz)=dt$. We put the way of how to derive dual forms in the Appendix A.

\section{Mathematical Preliminaries: Newman-Penrose Formalism}

\subsection{Newman-Penrose Formalism}
In this section we shall learn how to derive the components of Ricci and Weyl tensor by the Newman-Penrose (NP) formalsim.  The NP formalism is a mathematical framework used for analyzing the gravitational field. See Carmeli \cite{carmeli2001classical}.
The most powerful and economical work of NP formalism is the introduction of a complex null tetrad. A null tetrad is a set of four mutually orthogonal vectors at each point in spacetime. These null tetrads are chosen to be null and form a basis, that is, the norm=0 or $A_\mu A^\mu=0$. The complex nature of the tetrad allows for the separation of spacetime into two sets of two complex 1-forms.
Using the null tetrad, various complex scalars known as spin coefficients are defined.
These coefficients are constructed from contractions of the Riemann curvature tensor and represent different components of the gravitational field.

We therefore introduce the 4-null vectors. They are 2 real $l_\mu$, $n_\mu$ and 2 complex $m_\mu$, $\bar{m_\mu}$, where $\bar{m}_\mu$ is complex conjugate of $m$. They follow the conditions that

\begin{equation}
  \begin{array}{lr}
    1) l_\mu l^\mu=l^2=0=n^2=m^2=\bar{m}^2.  \\
    2) l\cdot n=1=g_{\mu\nu}l^\mu n^\nu=g^{\mu\nu}l_\mu l_\nu, \; m\cdot\bar{m}=-1. \\
    $Note that this is for (+,-,-,-) and for (-,+,+,+) they pick up a minus sign after dot product.$ \\
    3) l\cdot m=n\cdot m=l\cdot \bar{m}=n\cdot \bar{m}=0.\\
  \end{array}
\end{equation}

Then we define that

\begin{equation}
  \begin{array}{lr}
    1) l=$Null basis 1-form$=l_\mu dx^\mu  \\
    2) n=n_\mu dx^\mu \\
    3) m=m_\mu dx^\mu.\\
  \end{array}
\end{equation}

After make the exterior derivative of the 1-forms, we may find 12 complex spin coefficients, $\alpha$, $\beta$, $\gamma$,... There are 24 components of them (12 in real parts and 12 in imaginary parts), corresponding to 24 potentials of $\Gamma^\mu_{\alpha\beta}$. Or we can call these spin coefficients the optical scalars. For example, $Re(\rho)$= expansion and converge, $Im (\rho)$=twist/rotation, and $|\sigma|$=shear. See figure 1.5. Thus, if we want to study the optics of gravitation, we must find all spin coefficients. A summary of NP formalism is given below in Appendix B.

\begin{figure}
  \centering
  \includegraphics[width=1\textwidth]{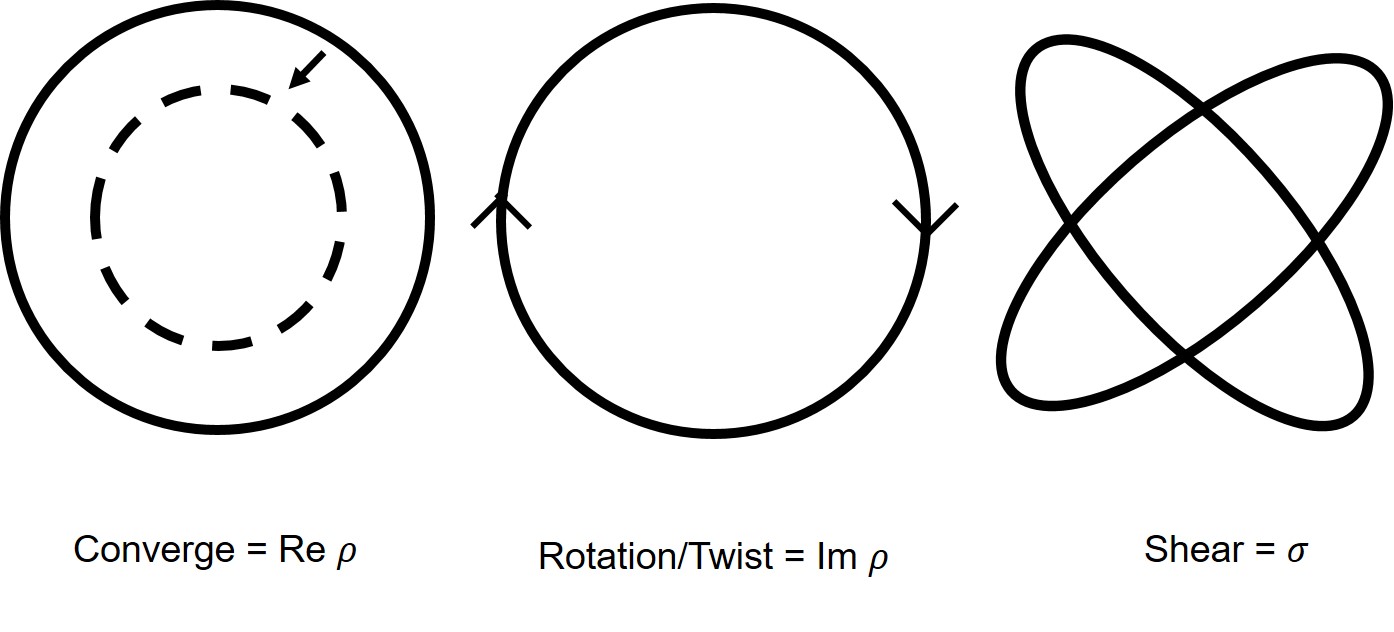}
  \caption{\centering This pictures show how circles of particles will be distorted with the passage of gravitational waves. For example, if $\rho$= Real, then particles will not undergo Rotation but converge. Also, $Im\; \rho$= rotation/twist and $\sigma$ the shear, etc.}\label{f2}
\end{figure}

And the dual forms of the tetrad reads

\begin{equation}
  \begin{array}{lr}
   *(l\wedge n)=im\wedge \bar{m},\;*(l\wedge m)=-il\wedge m,\;*(l\wedge \bar{m})=il\wedge \bar{m},\;\\
   *(n\wedge m)=in\wedge m,\;*(n\wedge \bar{m})=-in\wedge \bar{m}),\;*(m\wedge \bar{m})=il\wedge n,\\

   *l=il\wedge m\wedge \bar{m},\;*n=-in\wedge m\wedge \bar{m},\;*m=-il\wedge n\wedge m,\;*\bar{m}=il\wedge n\wedge \bar{m}.\;;
  \end{array}
\end{equation}

Therefore, one might be able to find $dF$ and $d*F$ via

\begin{equation}
  \begin{array}{lr}
    F=-(\Phi_1+\bar{\Phi}_1)l\wedge  n+\Phi_2l\wedge  m+\bar{\Phi}_2l \wedge \bar{m}-\bar{\Phi}_0n\wedge m-\Phi_0n\wedge \bar{m}+(\Phi_1-\bar{\Phi}_1)m\wedge \bar{m}\\

           \Phi_0=F_{\mu\nu}l^\mu m^\nu\\

         \Phi_1=\frac{1}{2}F_{\mu\nu}(l^\mu n^\nu+\bar{m}^\mu m^\nu)\\

         \Phi_2=F_{\mu\nu}\bar{m}^\mu n^\nu\\

         4\pi T_{\mu\nu}=F_{\mu\alpha}F_\nu^\alpha-\frac{1}{4}g_{\mu\nu}(F^2)
  \end{array}
\end{equation}

One can see that there are

\begin{equation}
  \begin{array}{lr}
    \Phi_{00}(R),\Phi_{01}(C),\Phi_{02}(C),\Phi_{11}(R),\Phi_{12}(C),\Phi_{22}(R),\to\; $Ricci components$ \\
    \Lambda\\
    \Psi_0(C),\Psi_1(C),\Psi_2(C),\Psi_3(C),\Psi_4(C) \to\;$Weyl scalars$,\\
  \end{array}
\end{equation}

where $R$ and $C$ are real and complex numbers respectively. Complex numbers means 2 components inside. Therefore, there are totally 20 components. Here are physical properties of them.

\begin{equation}
  \begin{array}{lr}
    \Lambda=\frac{R}{24},\;R$ is the curvature scalar$\\
    \Phi_{ij}\;$are the Ricci-NP scalars related to the energy-momentum tensor$\\
    \Psi_i\;$ are the Weyl scalars. They could be interpreted as the pure gravitational field$.\\
  \end{array}
\end{equation}

One can try an example to derive these NPs , Riccis, and Weyls by an example of $ds^2=2dudv-(F(v)dx)^2-(G(v)dy)^2$, we put the calculation in the Appendix C.

\subsection{Classification of Weyl Tensor}
Gravitational fields are usually classified according to the Petrov classification of the Weyl tensor. Please see \cite{petrov2016einstein}. This is based on the number
of its distinct principal null directions and the number of times these are
repeated. Let us consider the Debever-Penrose equation $k_{[\rho} C_{\alpha]\beta\gamma[\delta} k_{\sigma]}  k^\beta k^\gamma=0$, where $k^\alpha$ are 4 null vectors and $C_{\alpha\beta\gamma\delta}$ the Weyl tensor. To find the roots of the equation and all the possibilities are shown in  1.6. Type I is a general case that all $k^\alpha$ are different. Type II is one double and two simple roots. Type D has two double toots. This is a special case that Schwarzchild and Kerr spacetime are in this category. Type III has triple roots of $k^\alpha$. And in Type N, $k^\alpha$ are all the same. In Type D, a relation between Weyl scalars, $9\Psi_0\Psi_4=\Psi^2_2$, is of important.
\begin{figure}
  \centering
  \includegraphics[width=1\textwidth]{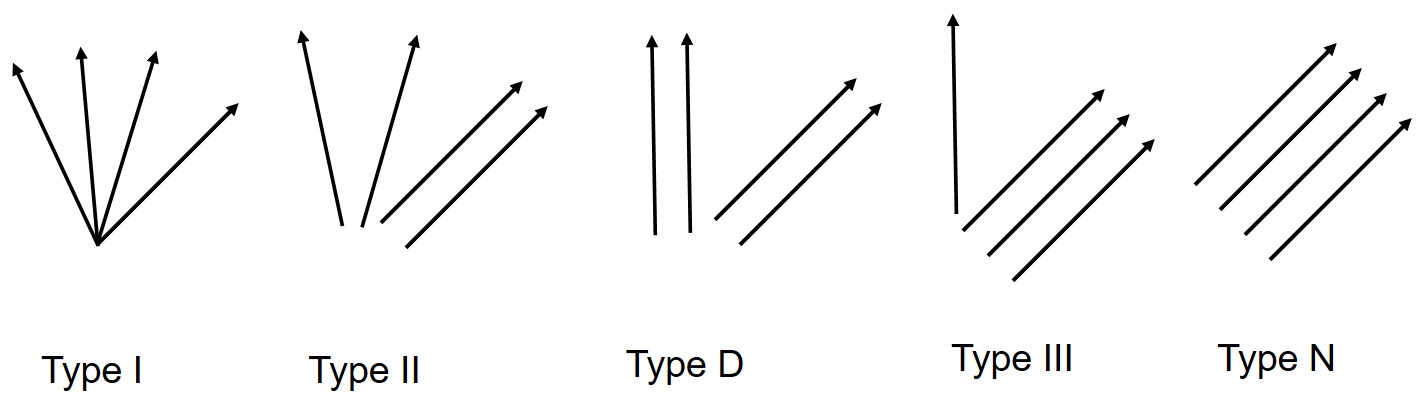}
  \caption{\centering The classification of 5 kinds of gravitational fields.}\label{f2}
\end{figure}

\section{Mathematical Preliminaries: Sign Conventions}
For different authors, the convention might be different. To fix this bothersome bump, we shall list the most probably $\pm$ sign conventions in some books. But we should stick to the idea that energy $\rho$ is always positive, then one can prospect the sign problem easier.

\begin{equation}
  \begin{array}{lr}
    \eta_{\mu\nu}=\eta^{\mu\nu}=[S_1]diag(-1,+1,+1,+1),$ Flat case$.\\
    R^\mu_{\alpha\beta\gamma}=[S_2](\Gamma^\mu_{\alpha\gamma,\beta}-\Gamma^\mu_{\alpha\beta,\gamma}
    +\Gamma^\mu_{\sigma\beta}\Gamma^\sigma_{\alpha\gamma}
    -\Gamma^\mu_{\sigma\gamma}\Gamma^\sigma_{\alpha\beta}),$ Riemann tensor$\\
    G_{\mu\nu}=[S_3]\frac{8\pi G}{c^4}T_{\mu\nu}, \;$(Einstein tensor)$=R_{\mu\nu}-\frac{R}{2}g_{\mu\nu}\\
    
    R_{\mu\nu}=$ Ricci tensor$, \; R=g^{\alpha\beta}R_{\alpha\beta}=$ scalar curvature$.\\
    R_{\mu\nu}=[S_2][S_3]R^\alpha_{\mu\alpha\nu}\\

  \end{array}
\end{equation}

\vspace{-10mm}
\begin{table}
    \caption{\centering Books with different signs.}
    \label{table:1}
    
    \begin{tabular}{|c c c c|}
     \hline
      & $S_1$ & $S_2$ & $S_3$ \\ [0.5ex]
     \hline\hline
     MTW (Misner, Thorne, Wheeler): & + & + & + \\
     S. Weinberg & + & - & - \\
     Rindler, Peacock, Narlikar, Padmanapan, Griffiths  & - & + & - \\
      [1ex]
     \hline
    \end{tabular}
\end{table}


\chapter{Colliding fields in two-dimensional dilaton gravity background}\label{ch:preliminary}

The 1+1-Dimensional gravity grasps the very essential physics of understanding how the waves colliding without affected by other dimensions. In this thesis, we shall choose the CGHS model as our start point, then go further to more complicated cases, especially the creation of wormholes by ghost fields.

In the CGHS model, the double-null coordinates $(u, v)$ with the element would be

\begin{equation}
  ds^2=-2e^{2\phi}dudv=-2Fdudv.
\end{equation}

Again, the CGHS Lagrangian reads

\begin{equation}
  S=\int d^2x\sqrt{-g}[e^{-2\phi}(R+4(\nabla\phi)^2+4\lambda^2-\frac{1}{2}(\nabla f)^2+\frac{1}{2}(\nabla g)^2],
\end{equation}
where the function $f$ is the real field but the $g$ field will pick up an extra i factor, therefore, $g$ should be recognized as the ghost field. From the metric 2.1, the scalar curvature is $\frac{2}{F^3}(FF_{uv}-F_uF_v)$, that is, $R=4e^{-2\phi}\phi_{uv}$, where the sub $u/v$ implies partial derivatives. Note, we choose $F=e^{2\phi}$ from the outset, which will make $\sqrt{-g}$ cancel the $e^{-2\phi}$ in the Lagrangian. This trick solves the differential equations analytically. Furthermore. we define $r=2e^{-2\phi}$, which plays the role of radius. Then we apply the
variational principle and integrate by parts twice. For example, $I=\int dudv FG_{uv}$, the method should be

\begin{equation}
  \begin{array}{lr}
    \delta I=0 \rightarrow\int dudv [\delta FG_{uv}+F(\delta G)_{uv}]=0\\
     $Firstly, $G_{uv}=0\\
    $Then, we make integration by parts on $ F(\delta G)_{uv}\\
    =[F(\delta G)_{u}]_v-F_v(\delta G)_{u}, $the first term vanishes because of surface$.\\
    $And another by parts brings us$\\
    -[F_v(\delta G)]_u+F_{uv}(\delta G), $the first term vanishes because of surface$.\\
    \therefore F_{uv}=0

  \end{array}
\end{equation}

Therefore, applying the variational principle to 2.2 brings us the differential equations of motion

\begin{eqnarray}
f_{uv}=0=g_{uv},
\\
 r_{uv}=-4\lambda^2,
\\
r_{uu}=g_{u}^2-f_{u}^2,
\\
r_{vv}=g_{v}^2-f_{v}^2.
\end{eqnarray}

In the dilaton vacuum case $(f=g=0)$, the forgoing set of equations integrate as

\begin{equation}
 r(u, v)=2m-4\lambda^2uv,
\end{equation}

where $m>0$ is a positive constant that plays the role of mass. The energy expression \cite{hayward2002dilatonic}

\begin{equation}
 E=\frac{r}{2}(1-\frac{(\bigtriangledown r )^2}{4\lambda^2r^2}),
\end{equation}

gives $E=m$, justifying the interpretation of m as the mass. The scalar curvature gives

\begin{equation}
 R=\frac{4m\lambda^2}{m-2\lambda^2uv},
\end{equation}

which shows that the hyperbolic arc $uv=\frac{m}{2\lambda^2} $ is a singularity. The case $m=0$ is obviously a flat space that will not be considered. In order to see the black hole/wormhole structure of the foregoing solution, we transform the line element with the solution (2.8) into a more familiar form. The line element (2.1) reads

\begin{equation}
 ds^2=-\frac{4}{r}dudv,
\end{equation}

where $r(u, v)$ is given by (2.8). From the null coordinates, we transform (2.11) into the ‘Schwarzchild-type’ coordinate $(t, r)$ as follows

\begin{enumerate}
    \item For $r>2m$, we choose $(u>0, v<0)$ as
    \begin{equation}
    \begin{split}
    &2\lambda u=\sqrt{r-2m}e^{t/2} \\
    &2\lambda v=-\sqrt{r-2m}e^{-t/2}
    \end{split}
    \end{equation}

    \item For $r<2m$, with $(u>0, v>0)$ we make the choice
    \begin{equation}
    \begin{split}
    &2\lambda u=\sqrt{2m-r}e^{t/2} \\
    &2\lambda v=\sqrt{2m-r}e^{-t/2}
    \end{split}
    \end{equation}
\end{enumerate}

Both cases lead to the same form of the line element

\begin{equation}
  ds^2=\frac{1}{4\lambda^2}[-(1-\frac{2m}{r})dt^2+\frac{dr^2}{r(r-2m)}],
\end{equation}

in which $\lambda$ is clearly seen to be an indispensable parameter. Recalling the role played by an independent existence of a cosmological constant, we judge that $\lambda^2$ represents the cosmological constant provided $m\ne0$. For $m=0$, we have from (2.14) that it becomes a conformally flat spacetime without a cosmological constant. We remark that the inverse process of the above transformation from $(t, r)$ to $(u, v)$ in 3+1-dimensions was introduced first by Kruskal \cite{kruskal1960maximal} and Szekeres \cite{szekeres1960singularities,szekeres1970colliding}. Once our dilatonic metric is cast into this form, we apply the energy definition of Misner and Sharp \cite{misner1964relativistic}, expressed in the form

\begin{equation}
 E=\frac{r}{2}(1-g^{\mu\nu}r_{,\mu}r_{,\nu})
\end{equation}
Since the second term in the parenthesis vanishes, for the line element (2.14), it represents at $r=2m$, the horizon of a black hole or the mouth of a wormhole. We obtain from (2.15), that $E=m$.

\section{Colliding Waves in the CGHS Spacetime}

In this section we consider the collision of various fields the 1+1-dimensional spacetime of the dilaton vacuum \cite{halilsoy2023colliding}.

\subsection{Collision of Bounded Real Fields}
We consider the real field only and being chosen by

\begin{equation}
 f(u, v)=-1+\cos^2(au\theta(u))+\cos^2(bv\theta(v)),
\end{equation}

with ($a$,$b$) constants and step function $\theta$. 

The reason for such a choice and also the wave profiles in the following sections, is the symmetry $u\leftrightarrow v$, simple boundary conditions at the boundaries and the boundedness of the functions. For ($u<0$, $v<0$), we have the simplest form for the field, namely $f=1$, which is equivalent to no field, leaving only the CGHS background. We have already explained in the introduction the advantages of the $cos( )$ function as an incoming profile. It is a bounded function and it doesn’t create a delta function in the second derivative.We recall that in the problem of colliding electromagnetic waves in general relativity also similar problems were overcome by such $cos( )$ ansatz \cite{griffiths2016colliding,tahamtan2016colliding,bell1974interacting,halilsoy1988colliding}. Once we choose $f(u, v)$, we solve the field equation (2.4) and (2.5) in the form

\begin{equation}
 r(u, v)=A(u)+B(v)-4\lambda^2uv,
\end{equation}
where $A(u)$ and $B(v)$ depend on the different null coordinates and satisfy

\begin{eqnarray}
A_{uu}=-f_u^2,
\\
 B_{vv}=-f_v^2.
\end{eqnarray}

Upon integration , by choosing the integration constants appropriately, we obtain

\begin{equation}
\begin{split}
r(u, v)=2m-4\lambda^2uv+
 \frac{1}{32}(2-\cos(4au\theta(u)-\cos(4bv\theta(v))\\
 -
 \frac{1}{4}(a^2u^2\theta(u)+b^2v^2\theta(v)).
\end{split}
\end{equation}

We recall that in the term, $-4\lambda^2uv$, which is from the CGHS background, the null coordinates do not have step functions. The problem now can be formulated as a collision in the following form $(Fig(2.1))$.

\subsubsection{Region $I$, $(u<0, v<0)$}

\begin{equation}
\begin{split}
\\
   f(u, v)=+1\nonumber,\\
   r(u, v)=2m-4\lambda^2uv. \nonumber
   \\
   \\
\end{split} 
\end{equation}

\subsubsection{Region $II$, $(u>0, v<0)$}
\begin{equation}
\begin{split}
\\
    f(u)=\cos^2(au),\nonumber\\
    r(u, v)=2m-4\lambda^2uv+\frac{1}{32}(1-\cos(4au))-
 \frac{1}{4}a^2u^2. \nonumber 
   \\
   \\
\end{split} 
\end{equation}

\subsubsection{Region $III$, $(u<0, v>0)$}

\begin{equation}
\begin{split}
\\
    f(v)=\cos^2(bv),\nonumber\\
    r(u, v)=2m-4\lambda^2uv+\frac{1}{32}(1-\cos(4bv))-
 \frac{1}{4}b^2v^2.\nonumber  
 \\
 \\
\end{split}
\end{equation}

\subsubsection{Region $IV$, $(u>0, v>0)$}

\begin{equation}
\begin{split}
\\
f(u, v)=-1+\cos^2(au)+\cos^2(bv),\nonumber\\
r(u, v)=2m-4\lambda^2uv
  +\frac{1}{32}(2-\cos(4au)-\cos(4bv))\\
  -\frac{1}{4}(a^2u^2+b^2v^2).\nonumber
  \\
  \\
\end{split}
\end{equation}

All the foregoing expressions are exact, satisfying the boundary conditions at $u=v=0$. In analogy with the colliding gravitational shock waves \cite{szekeres1970colliding}, we can give a perturbative expression. To the leading orders, we obtain for $|u|<<1$ and $|v|<<1$,

\begin{equation}
 f(u, v)\approx1-a^2u^2-b^2v^2+\frac{1}{3}(a^4u^4+b^4v^4)+...
\end{equation}

\begin{equation}
 r(u, v)\approx2m-4\lambda^2uv-\frac{1}{3}(a^4u^4+b^4v^4)+...
\end{equation}

There is no need to state that to the orders $O(u^2)$ and $O(v^2)$, this is same as the dilatonic black hole. We note that the incoming functions were bounded in this case. If we considered instead the unbounded case for the function $f(u, v)$, given by $f(u, v)=a^2u^2\theta(u)+b^2v^2\theta(v)$, we obtain $r(u, v)$ exactly as

\begin{equation}
  r(u, v)=2m-4\lambda^2uv-\frac{1}{3}(a^4u^4+b^4v^4)
\end{equation}

which coincides with the approximate form (2.22). The scalar curvature corresponding to this solution is given by

\begin{equation}
  R=\frac{4}{r}[\lambda^2(2m+a^4u^4+b^4v^4)+\frac{4}{9}a^4b^4u^3v^3]
\end{equation}

in which $r$ is given in (2.23) and for $r=0$, it gives a naked singularity.

\subsection{Collision of Hyperbolic Ghost Fields}

For this case, we choose the real field $f=0$, and the ghost field as

\begin{equation}
 g(u, v)=-1+\cosh^2(au\theta(u))+\cosh^2(bv\theta(v)),
\end{equation}

in which $(a, b)$ are again arbitrary constants. In analogy with the section (2.1.1), we integrate $r(u, v)$ with the appropriate integration constants to obtain for $(u>0, v>0)$

\begin{equation}
\begin{split}
r(u, v)=2m-4\lambda^2uv-\frac{1}{16}\\
  +\frac{1}{32}(\cosh(4au)+\cosh(4bv)) -\frac{1}{4}(a^2u^2+b^2v^2).
\end{split}
\end{equation}

Up to certain signs, and the unbounded hyperbolic functions instead of bounded trigonometric ones, colliding ghosts has similarities with the real fields. A similar figure to $Fig.(2.1)$ can be plotted where $f$ is replaced by $g$. In the interaction region (Region $IV$), expansion to the lowest orders $|u|<<1$ and $|v|<<1$ gives

\begin{equation}
 g(u, v)\approx1+a^2u^2+b^2v^2+\frac{1}{3}(a^4u^4+b^4v^4)+...
\end{equation}

\begin{equation}
 r(u, v)\approx2m-4\lambda^2uv+\frac{1}{3}(a^4u^4+b^4v^4)+...
\end{equation}

which  are comparable with the expansions of the colliding real fields. It is observed that to the order $O(u^2)$, $O(v^2)$ we have again a black hole in the interaction region. It should also be added that in case $g=a^2u^2\theta(u)+b^2v^2\theta(v)$, instead of the hyperbolic function one obtains $r(u, v)$ exactly as

\begin{equation}
  r(u, v)=2m-4\lambda^2uv+\frac{1}{3}(a^4u^4+b^4v^4)
\end{equation}

The change of sign in the last term in this expression and that of (2.23) reflects the difference between the real and ghost fields, respectively.

\begin{figure}
  \centering
  \includegraphics[width=0.8\textwidth]{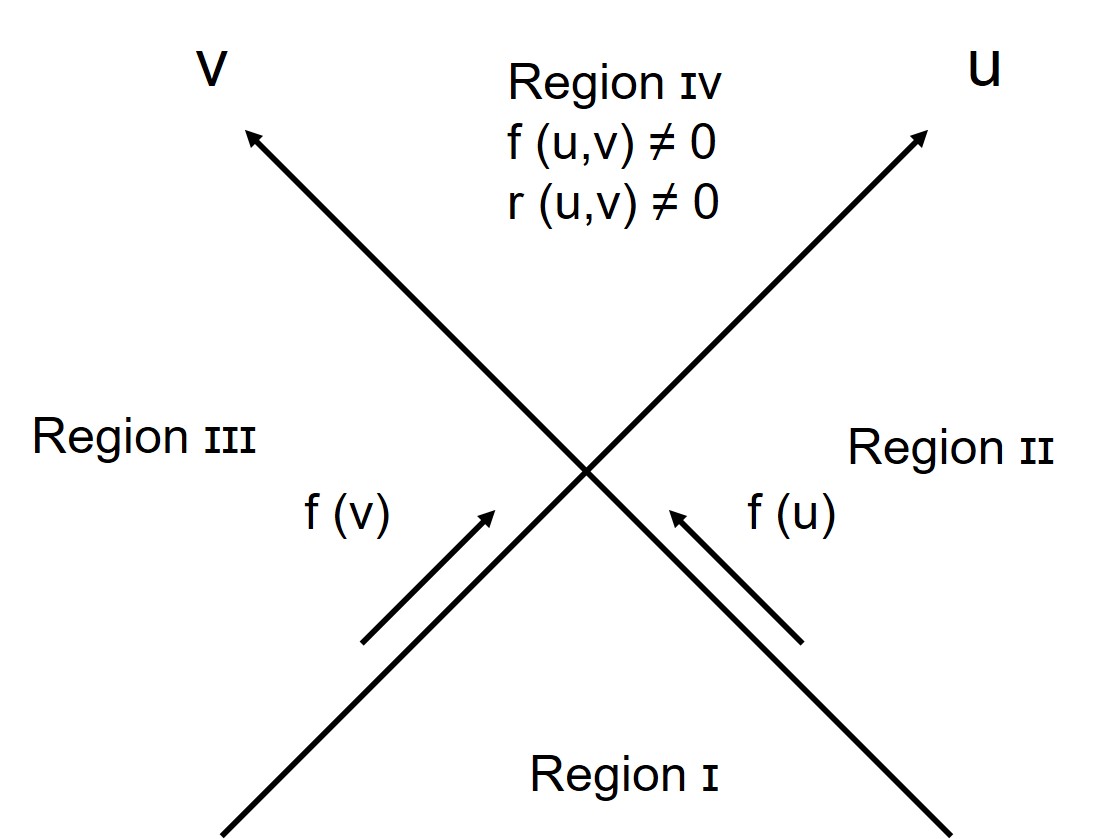}
  \caption{\centering The picture describing the collision of two real fields. Region $I$ contains only a background dilaton, $r=2m-4\lambda^2uv$, with a trivial field $f=1$. Region $II$, has the incoming field $f(u)=\cos^2(au)$, which distorts also the background as given in the text. Region $III$, is the symmetric region of $II$, with $u\leftrightarrow v$ and $a\leftrightarrow b$. Region $IV$ is the interaction region having $f(u, v)$ and $r(u, v)$ with maximum terms that reduce to the appropriate incoming expressions. }\label{f1}
\end{figure}

\subsection{Collision of Real and Ghost Fields}

With reference to Fig.2.1, we consider an incoming real field in Region $II$ given by

\begin{equation}
 f(u)=\cos^2(au\theta(u))
\end{equation}

and a ghost field in Region $III$ described by

\begin{equation}
 g(v)=\cosh^2(bv\theta(v)).
\end{equation}

These two fields interact at $u=v=0$ and develop into the Region $IV$, where the function $r(u, v)$ is given by

\begin{equation}
\begin{split}
 r(u, v)=2m-4\lambda^2uv+\frac{1}{32}(\cosh(4bv\theta(v))-\cos(4au\theta(u)))\\
 -\frac{1}{4}(a^2u^2\theta(u)+b^2v^2\theta(v)).
\end{split}
\end{equation}

It can be checked that this satisfies all boundary conditions and in particular the respective ghost and real field conditions

\begin{equation}
 r_{uu}=-f_u^2,
\end{equation}

\begin{equation}
 r_{vv}=g_v^2,
\end{equation}
are satisfied by the foregoing expressions. An expression of $cosh()$ and $cos()$ functions reveal that to the second order we have

\begin{equation}
 r(u, v)\approx 2m-4\lambda^2uv+O(u^4, v^4),
\end{equation}
which is the black hole condition. In general, however, the expression (2.29) involves more complicated terms to argue in favor of a black hole.

\subsection{Collision of Two Linear Ghost Fields}

This type of ghost fields was already considered in \cite{hayward2002dilatonic}. We choose it here to discuss the problem of collision of such ghosts. We take

\begin{equation}
 g(u, v)=2\lambda(u\theta(u)-v\theta(v)),
\end{equation}
which implies that in the incoming regions we have linear fields with special coefficients. The field equations to be integrated are

\begin{eqnarray}
 r_{uv}=-4\lambda^2,
\\
r_{uu}=4\lambda^2\theta(u),
\\
r_{vv}=4\lambda^2\theta(v).
\end{eqnarray}

The integral for $r(u, v)$ is given in general by

\begin{equation}
 r(u, v)=2m-4\lambda^2uv+2\lambda^2(u^2\theta(u)+v^2\theta(v))+c_1u+c_2v,
\end{equation}
where $c_1$ and $c_2$ are arbitrary integration constants. We make choice for $c_1$ and $c_2$ and obtain the following two cases

\subsubsection{The Choice of $c_1=c_2=0$, for $u>0, v>0$.}
This leads to

\begin{equation}
 r(u, v)=2m+2\lambda^2(u-v)^2,
\end{equation}
and by defining new coordinates

\begin{equation}
\sqrt{2}u=t+z,
\sqrt{2}v=t-z,
\end{equation}
we obtain

\begin{equation}
 r(z)=2m+4\lambda^2z^2,
\end{equation}
which is a wormhole solution with the throat $r=2m$ at $z=0$. The case $z>0$ and $z<0$ are the two mirror images of the wormhole. Thus, as a result of two colliding linear ghosts we obtain a wormhole.

\subsubsection{The Choice of $c_1=c_2=-4\sqrt{m\lambda}$, for $u>0, v>0$.}
By inserting these constants into the integral (2.30), we obtain

\begin{equation}
 r(u, v)=2(\sqrt{m}-\lambda(u-v))^2,
\end{equation}

which gives the dilaton

\begin{equation}
 e^{-2\phi}=(\sqrt{m}-\lambda(u-v))^2,
\end{equation}

The line element reads now

\begin{equation}
 ds^2=\frac{-dt^2+dz^2}{(\sqrt{m}-\sqrt{2}\lambda z)^2},
\end{equation}

with the Ricci scalar $R=12\lambda^2$, which is regular. The two-dimensional dilation space can be expressed in the form

\begin{equation}
 ds^2=-e^{2\sqrt{2}\lambda Z}dt^2+dZ^2,
\end{equation}

where the new variable $Z(z)$ is defined by

\begin{equation}
Z(z)=-\frac{1}{\sqrt{2}\lambda}\ln(\sqrt{m}-\sqrt{2}\lambda z),
\end{equation}

which is again a wormhole solution in the new variable. It is seen that from both choices of the constants for $c_1$ and $c_2$, we obtain wormhole solutions as a result of colliding linear ghosts.

\subsection{Collapse of the Wormhole}

In the previous section, we described how a linear wormhole forms from the collision of two ghost fields. It can be easily checked that the fields

\begin{equation}
 g(u, v)=2\lambda(u\theta(-u)-v\theta(-v)),
\end{equation}

\begin{equation}
 r(u, v)=2m+2\lambda^2(u^2\theta(-u)+v^2\theta(-v)-2uv),
\end{equation}

also solves the field equations. We note that the present work differs from that of \cite{hayward2002dilatonic} by insertion of the step functions which makes the collision problem possible. As described in $Fig (2.2)$, the two ghosts collide first at $A$, forming a wormhole between $A$ and $O$. At $O$ the ghost fields vanish and this enforce the wormhole to collapse to a black hole. For completeness, let us add that between $A$ and $O$, a linear coordinate transformation on the null coordinate is required, which is out of our interest.


\section{Summary}
This study revisits the two-dimensional dilaton gravity model of CGHS,
In such a background that we are able to make the metric function equal to the dilaton, the collision of real and ghost fields in various combinations is explored. This work aims to address the lack of detailed analysis regarding field collisions in previous study \cite{hayward2002dilatonic}. Even though the problem is simplified to particles moving at light speed in one dimension, it remains not so easy. The dilaton, constructed from the cosmological constant, acts as a catalyst in these processes.

Results indicate that ghost fields create wormholes, while real fields primarily lead to naked singularities. However, collision ultimately leads to black hole formation to quadratic orders in $u$ and $v$. The addition of more fields and wave packets can yield an infinite class of solutions akin to gravitational waves \cite{szekeres1970colliding}. Notably, the collision of two ghost fields produces the most interesting result, reminiscent of the very first concept of a wormhole as a geometric model of an elementary particle \cite{einstein1935particle}. This was due to the minimum radius which characterizes a particle. Based on the dynamical equations we have shown that a wormhole radiates out two linear ghosts, which made the wormhole. These processes all take place in 1+1-dimensional CGHS model and to what extent such processes take place in higher dimensions remains to be seen. All these solid findings suggest truly that 1+1 is more than 2!

\begin{figure}
  \centering
  \includegraphics[width=0.8\textwidth]{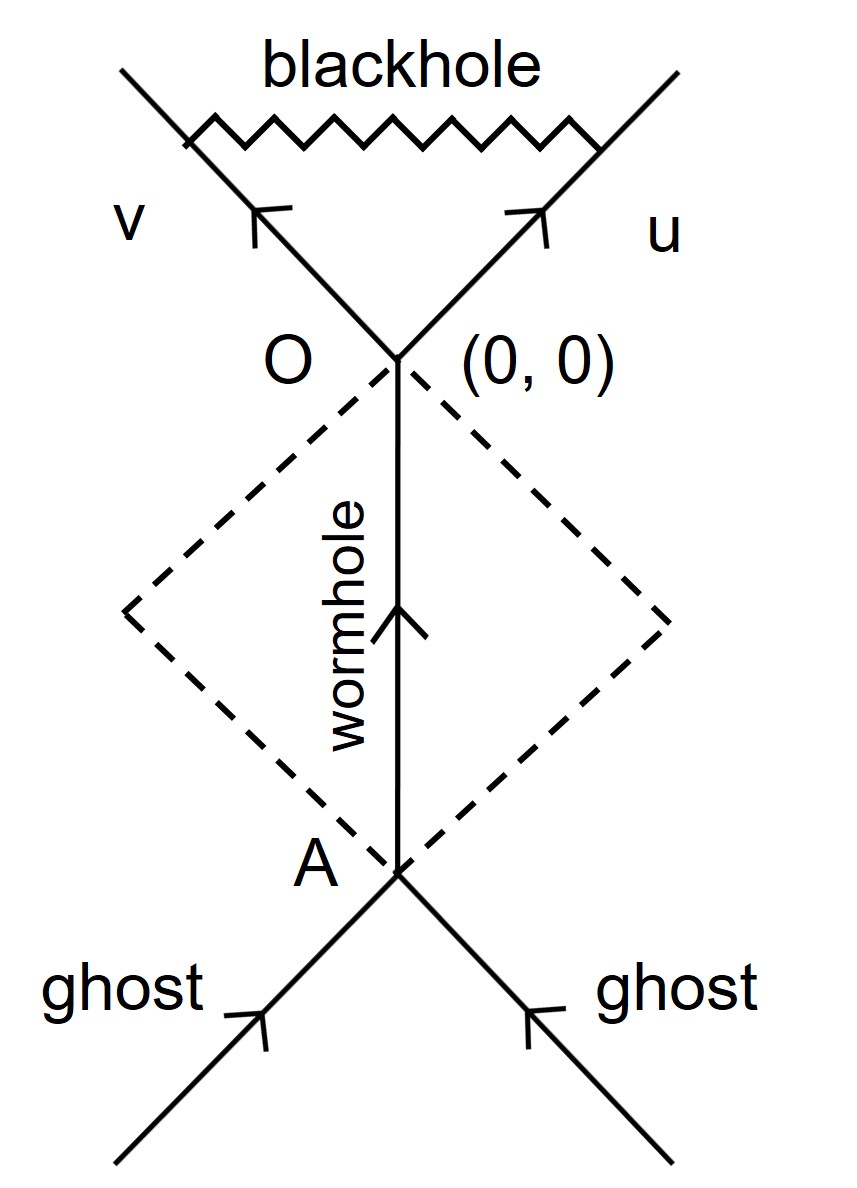}
  \caption{\centering Diagram for creation and collapse of a wormhole. The line $AO$ represents a wormhole with throat $r=2m$, for, $u<0$, $v<0$ with $r=2m+2\lambda^2(u-v)^2$ as a yield of colliding ghosts in the past at $A$. At point $O$ the ghosts cease to support the wormhole and as a result the wormhole collapses to the black hole $r=2m-4\lambda^2uv$, once again.}\label{f2}
\end{figure}

\chapter{Colliding Null Matter in a Model of Bumblebee Gravity}\label{ch:preliminary}
Bumblebee gravity or equivalently the global monopole and string dust, provide the very crucial parameter k (or the breaking constant $l$) to grasp the geometry of Schwarzchild-like black hole. We develop a toy model trying to depict the astrophysical jet erupted from accretion disk by this geometry and confine it by the parameter k. In other words, we provide a model to test the bumblebee gravity and understand the level of Lorentz symmetry breaking.

\section{Introduction of Bumblebee Gravity Geometry and Our Model}

Bumblebee gravity is a theoretical framework that extends general relativity by introducing an additional vector field known as the bumblebee field, which can lead to Lorentz symmetry breaking by introducing preferred directions in spacetime.
We shall introduce the mechanism of Bumblebee gravity in this section. Firstly, one may know the concept of symmetry breaking. For example, we can consider the Gibbs free energy \cite{peskin2018introduction}

\begin{equation}
    G(M)=a\cdot(T-T_c)M^2+b\cdot(T)M^4,
\end{equation}
where M is the magnetization and $T_c$ can be regarded as the Curie temperature. Symmetry breaks when temperature drops below the Curie temperature. We can see from the figure 3.1 that in 2 dimension, the magnetization will orient to either,say, right or left. If $b\cdot(T)=c$,

\begin{equation}
M =
\left\{
    \begin{array}{lr}
        \pm[\frac{a}{2c}(T_c-T)^\frac{1}{2}], \qquad   &  T < T_c\\
        0, \qquad    &  T \geq T_c.
    \end{array}
    \right\}
\end{equation}

The average value would be $<M>\ne0$ after symmetry breaking. Now, We can also consider the U(1) symmetry case, much like the magnetization projects into 2 dimension, as shown in figure 3.2. The potential around the trough=0.

\begin{figure}
  \centering
  \includegraphics[width=0.8\textwidth]{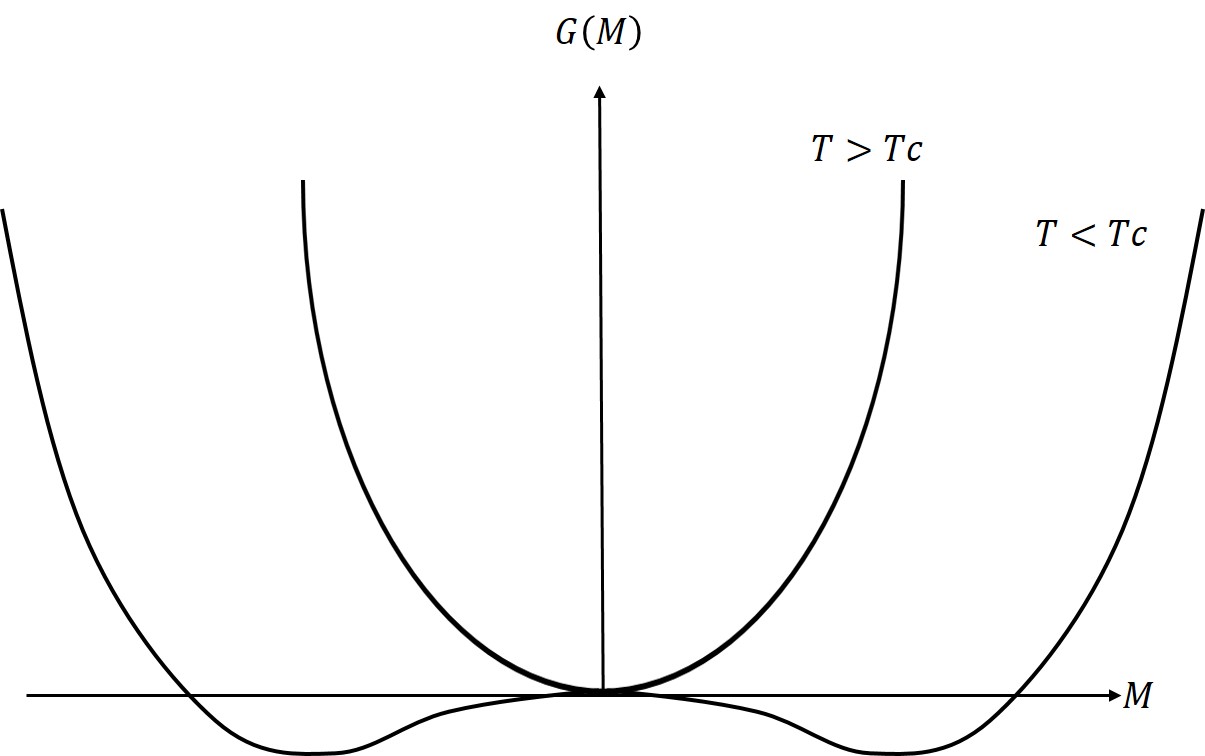}
  \caption{\centering The Gibbs free energy $G(M)$ at
temperatures above and below the critical temperature.}\label{f2}
\end{figure}

\begin{figure}
  \centering
  \includegraphics[width=0.8\textwidth]{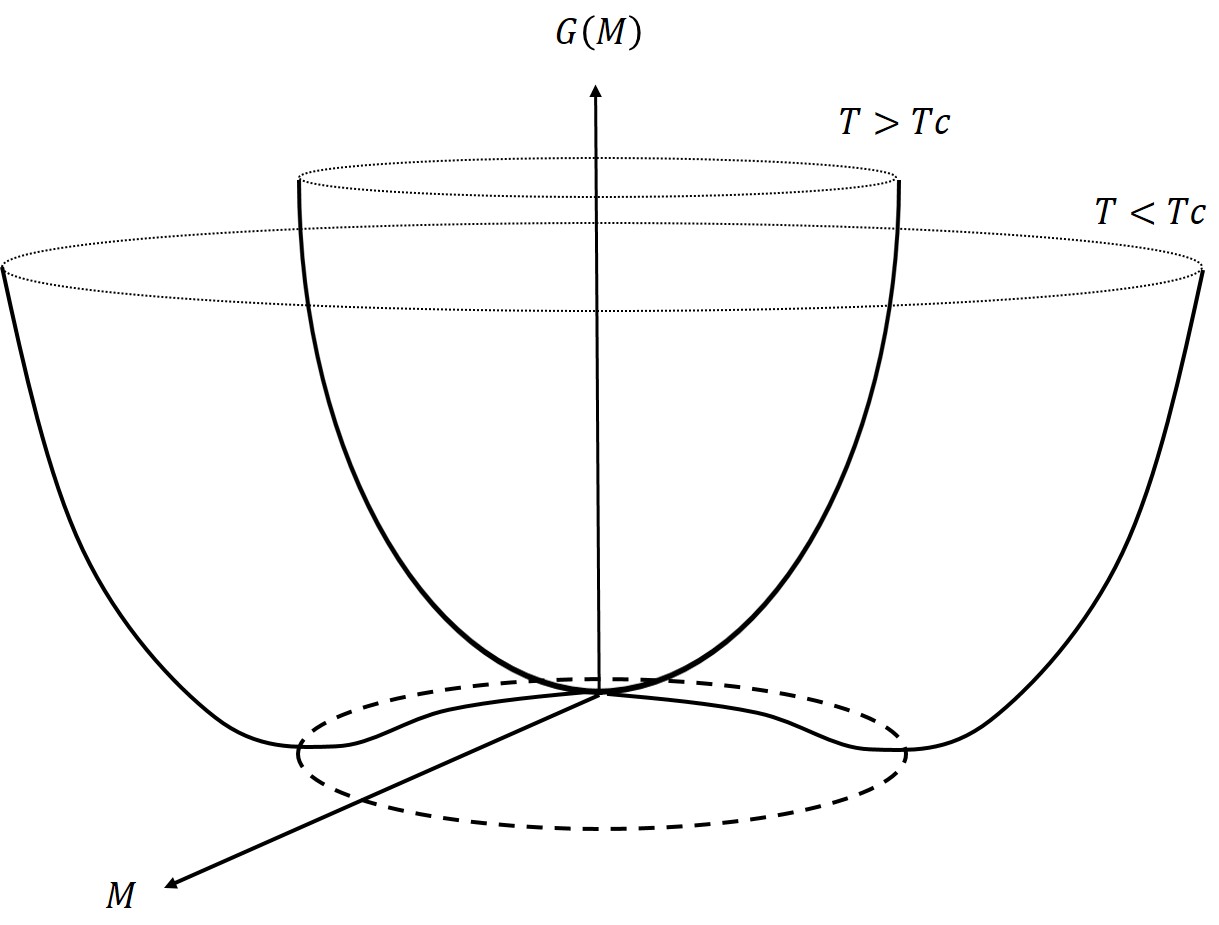}
  \caption{\centering Due to the potential instability of the origin point, particles would drop to the trough.}\label{f2}
\end{figure}

The concept of bumblebee gravity is to replace M by bumblebee field $B^\mu$. So that after symmetry breaking, we have

\begin{equation}
    <B^\mu>=b^\mu.
\end{equation}
In such way, we find that the potential

\begin{equation}
    V=V(B^\mu B_\mu\pm b^\mu b_\mu)=0,
\end{equation}
and its derivative also vanishes.

With these gears, we can consider the Lagrangian of $B^
\mu$ coupling to gravity,

\begin{equation}
    S=\int d^4x\; L_B=\int d^4x\; (L_g+L_{gB}+L_K+L_V+L_M),
\end{equation}
where $L_g$ is the pure gravitational
Einstein-Hilbert term, $L_{gB}$ the gravity-bumblebee coupling, $L_K$ the bumblebee kinetic and any self-interaction terms, $L_V$ the potential and $L_M$ the matter and others field contents and their couplings to the bumblebee field.
We can write the Lagrangian as

\begin{equation}
    L_B=\frac{e}{2\kappa}R+\frac{e}{2\kappa}\xi B^\mu B^\nu R_{\mu\nu}-\frac{1}{4}eB_{\mu\nu}B^{\mu\nu}-eV(B^\mu)+L_M,
\end{equation}

with $e=\sqrt{-g}$ , $\kappa=8\pi G$ and $\xi$ the coupling constant.

After the symmetry breaking, we may bring conditions 3.3 and 3.4 into the Lagrangian and consider 

\begin{equation}
    T_{\mu\nu}=\frac{-1}{\sqrt{-g}}\frac{\delta}{\delta g_{\mu\nu}}\sqrt{-g}L_B.
\end{equation}

We shall have $T_{\mu\nu}=T^M_{\mu\nu}+T^B_{\mu\nu}$, the matter part $T^M_{\mu\nu}$ and the bumblebee part $T^B_{\mu\nu}$ of energy-momentum tensor, respectively. Connecting the energy-momentum tensor to the Einstein equation

\begin{equation}
G_{\mu\nu}=R_{\mu\nu}-\frac{1}{2}Rg_{\mu\nu}=T_{\mu\nu},
\end{equation}

leads to

\begin{equation}
    R_{\mu\nu}=\kappa(T^M_{\mu\nu}-\frac{1}{2}g_{\mu\nu}T^M)
    +\frac{\xi}{2}g_{\mu\nu}\nabla_\alpha\nabla_\beta (B^\alpha B^\beta)+\frac{\xi}{4}g_{\mu\nu}\nabla^2 (B_\alpha B^\alpha).
\end{equation}

For a vacuum solution, $T^M_{\mu\nu}=0$. In order to solve the spherically symmetric vacuum solution, one should consider the Birkhoff metric 

\begin{equation}
    g_{\mu\nu}=diag(-e^{2\gamma},e^{2\rho},r^2,r^2sin^2\theta).
\end{equation}

and the ansatz of
the bumblebee field setting to $B_\mu=b_\mu=(0,0,b_r(r),0)=(0,0,|b|e^\rho,0)$ and $b^\mu b_\mu=b^2=const$. We then have \cite{casana2018exact}  

\begin{equation}
  \begin{array}{lr}
    R_{tt}=(1+\frac{l}{2})e^{2(\gamma-\rho)}[\partial^2_r \gamma+(\partial_r\gamma)^2-\partial_r\gamma\partial_r\rho+\frac{2}{r}\partial_r\gamma]+\frac{l}{r}(\partial_r\gamma+\partial_r\rho)e^{2(\gamma-\rho)}\\

    R_{rr}=(1+\frac{3l}{2})[-\partial^2_r\gamma-(\partial_r\gamma)^2+\partial_r\gamma\partial_r\rho+\frac{2}{r}\partial_r\rho]\\

    R_{\theta\theta}=(1+l)e^{-2\rho)}[r(\partial_r\rho-\partial_r\gamma)-1]+1-l(\frac{1}{2}r^2e^{-2\rho}[-\partial^2_r\gamma-(\partial_r\gamma)^2+\partial_r\gamma\partial_r\rho+\frac{2}{r}\partial_r\rho]+1)\\

    R_{\phi\phi}=sin^2\theta e^{-2\rho}[r(\partial_r\rho-\partial_r\gamma)-1]+1,
  \end{array}
\end{equation}

with the Lorentz symmetry breaking constant $l=\xi b^2$.
Solving the equations in vacuum condition, $R_{ii}=0$, to get $\gamma$ and $\rho$,
one can derive the Schwarzchild-like metric

\begin{equation}
  ds^2=(1-\frac{2M}{r})dt^2-(1+l)\frac{dr^2}{(1-\frac{2M}{r})}-r^2(d\theta^2+\sin^2\theta d\phi^2),
\end{equation}
When the symmetry is restored, $b\rightarrow 0$, the vacuum solution of black hole is recovered.

On the other hand, we observe that the global monopole in static form is identical to (3.12). Since the global monopole metric reads 

\begin{equation}
  ds^2=(1-2k-\frac{2m}{r})dt^2-\frac{dr^2}{(1-2k-\frac{2m}{r})}-r^2(d\theta^2+\sin^2\theta d\phi^2),
\end{equation}

which represents the asymptotic form of a global monopole \cite{dadhich1998schwarzschild}. The constant parameter $k$ characterizes the strength of the monopole. For $k=0$, we recover the vacuum Schwarzchild metric. The energy-momentum tensor $T_\mu^\nu$, is given from the Einstein equations

\begin{equation}
G_\mu^\nu = T_\mu^\nu =
\begin{pmatrix}
\frac{-2k}{r^2} & 0 & 0 & 0 \\
0 & \frac{-2k}{r^2} & 0 & 0 \\
0 & 0 & 0 &0 \\
0 & 0 & 0 &0
\end{pmatrix}
\end{equation}

which suggests that the spacetime represents also a cloud of strings \cite{letelier1979clouds}. It can be easily checked that up to a scale transformation of time, we have in the foregoing metrics (3.12) and (3.13), the relations

\begin{equation}
  M=\frac{m}{1-2k},
\end{equation}

\begin{equation}
  l=\frac{2k}{1-2k},
\end{equation}
which make the two metrics identical. Once we show that these are identical, we choose (3.12) also as the representative of (3.13).

In general relativity, a static metric typically encompasses a dynamical region where, through a coordinate transformation, waves moving in opposite directions can be introduced. This dynamic region enables the definition of plane waves as substructures within the underlying spacetime and their subsequent collision. Notably, within the inner horizon region of a Schwarzschild black hole, the spacetime becomes dynamic, while the region outside the horizon remains static—a behavior observed across all Schwarzchild-like black hole solutions in various gravity theories.

We employ Rosen spacetimes \cite{rosen1937plane} of double-null coordinates with null source. A null source can be understood as a static source with an infinite boost, resulting in observers perceiving the source moving at the speed of light. Such null sources, often represented by null shells, have been extensively studied in the past within general relativity \cite{dray1986gravitation,wang1992gravitational,halilsoy2000comment}. Our collision model serves as a mathematical toy model for describing astrophysical jets observed at the central regions of strongly attractive sources, such as black holes. The collisions of neutral matter boosted to the speed of light produce impulsive and shock sources alongside gravitational waves.

By transforming the static metric (3.13) into double-null coordinates, we investigate the collision of boosted sources. This collision process yields new components of energy-momentum and results in impulsive and shock sources accompanied by gravitational waves. Under this circumstance, We try to describe the astrophysical jets, which are narrow streams of high-energy particles and electromagnetic radiation that are ejected from the vicinity of certain astronomical objects, such as black holes.

\section{The Method of Collision}

We apply a coordinate transformation on (3.13) to transform it into the double-null coordinate form so that we can formulate the problem of collision of sources \cite{halilsoy2024colliding}. By source clearly, it is to be understood the boosted form of the static metric in null-coordinates.

There is no loss of generality in making the choice for mass

\begin{equation}
  m=1-2k,
\end{equation}
and apply the coordinate transformation

\begin{equation}
  t=\frac{\sqrt{2}x}{1-2k},
\end{equation}

\begin{equation}
  \phi=\frac{\sqrt{2}y}{1-2k},
\end{equation}

\begin{equation}
  r=1+\sin(u+v),
\end{equation}

\begin{equation}
  \theta=\frac{1}{\sqrt{1-2k}}(u-v)-\frac{\pi}{2}.
\end{equation}

Upon rescaling
\begin{equation}
  ds^2\rightarrow\frac{1-2k}{2}ds^2,
\end{equation}
we obtain the line element in the form

\begin{equation}
  ds^2=X^2(2dudv-\cos^2(\frac{u-v}{\sqrt{1-2k}})dy^2)-\frac{\Delta}{X^2}dx^2,
\end{equation}
where we abbreviated

\begin{equation}
  X=1+\sin(u+v),
\end{equation}
and
\begin{equation}
  \Delta=\cos^2(u+v).
\end{equation}

It should also be added that under the transformation used the parameter $k$ is restricted as
\begin{equation}
0<k<\frac{1}{2},
\end{equation}
or for the corresponding Lorentz breaking factor

 \begin{equation}
l>0.
\end{equation}

Once we cast our static metric (3.13) (or (3.12)) into the double-null coordinate form, we are now ready to discuss the problem of collision.


\section{Collision of Null-matter Sources}

We insert into the global metric (3.23) the Penrose substitutions

 \begin{equation}
u\rightarrow u\theta(u),
v\rightarrow v\theta(v).
\end{equation}

We also make scaling in the null coordinates as

 \begin{equation}
u\rightarrow au,
v\rightarrow bv,
\end{equation}
where $(a, b)$ are constants with the units of inverse length in the geometrical units. In physical units $(a, b)$ measure energy and for that reason we use this freedom to incorporate the strength of energy of the incoming sources/waves. After the substitutions of (3.28) and (3.29) into our line element (3.23), we choose a null tetrad of Newman-Penrose (NP)\cite{newman1962approach} to compute the Ricci and curvature components. Our choices is

\begin{large}
\begin{equation}
    \begin{array}{lr}
         l_\mu=X(u, v)\delta_\mu^u,\\
         n_\mu=X(u, v)\delta_\mu^v,\\
         m_\mu=\frac{1}{\sqrt{2} X(u, v)}\cos(au\theta(u)+bv\theta(v))\delta_\mu^x \\
         +iX(u, v)\cos(\frac{au\theta(u)-bv\theta(v)}{\sqrt{1-2k}})\delta_\mu^y,
    \end{array}
\end{equation}
\end{large}

and the complex conjugate of $m_\mu$. We recall that our $X(u, v)$ is $X(u, v)=1+\sin(au\theta(u)+bv\theta(v))$.
In this choice of NP tetrad, the non-zero Ricci ($\phi_{AB}$) and Weyl components $\psi_A$ are tabulated in Appendix D. By inspecting these components, we can locate the singularities in the metric. We formulate first the problem of collision as follows. We divide the spacetime into four regions as figure 3.3.

\begin{figure}
  \centering
  \includegraphics[width=0.5\textwidth]{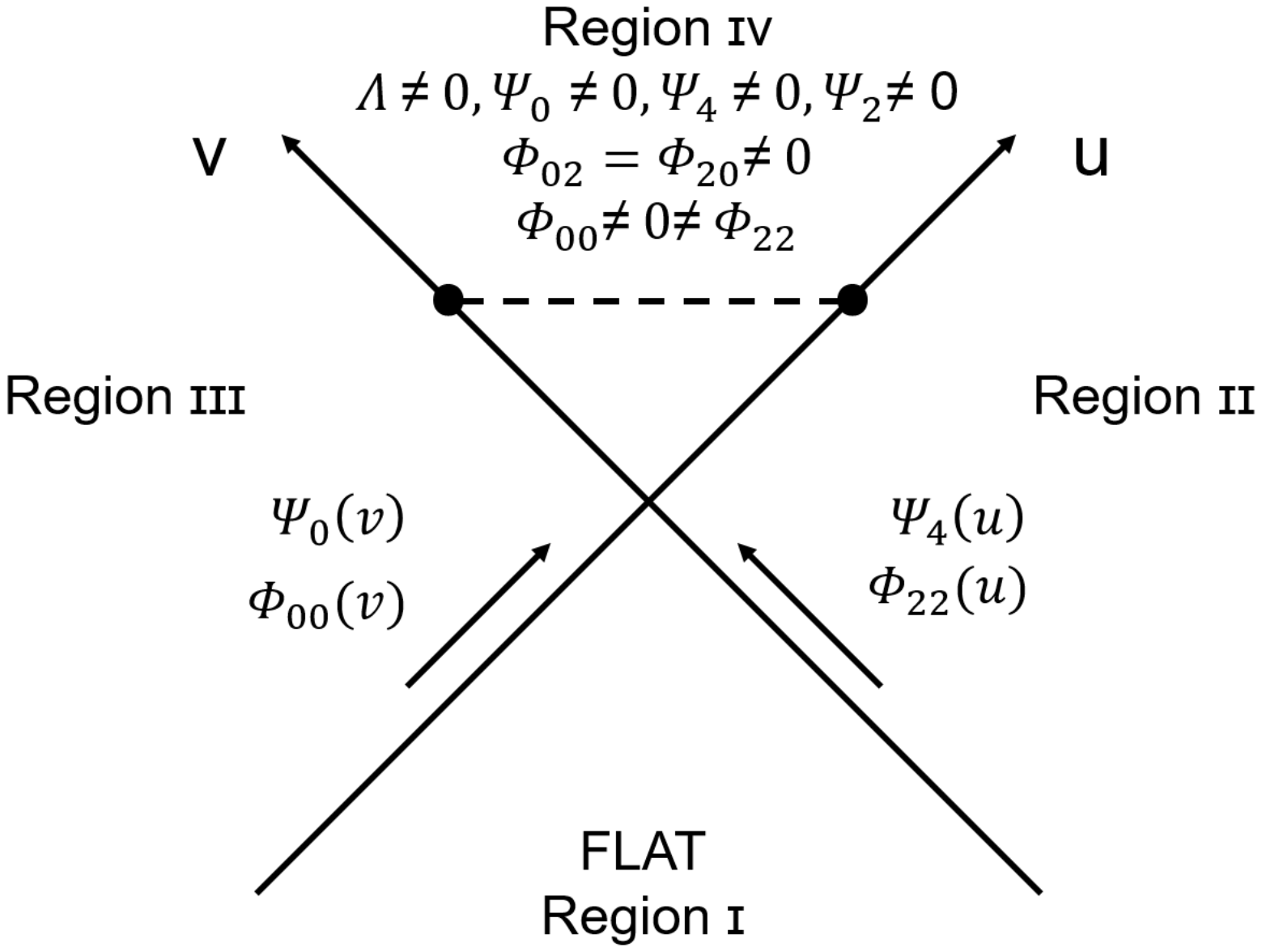}
  \caption{\centering Collision of sources $\Phi_{22}(u)$ and $\Phi_{00}(v)$ coupled with gravitational waves. There are null singularities at ($u=0, v=\frac{\pi}{2b\sqrt{1-2k}}$) and ($v=0, u=\frac{\pi}{2a\sqrt{1-2k}}$), shown as heavy dots on the null coordinates, where $\Psi_4$ and $\Psi_0$ diverge. The sources $\phi_{22}(u)$ and $\phi_{00}(v)$ also become divergent at these points. For the interaction region with $u>0, v>0$, however, the delta function $\delta(u)$ and $\delta(v)$ are not effective and we search for the other terms. We note that in the interaction region, we must impose the condition $au+bv<\pi\sqrt{1-2k}$, which otherwise confronts with the spacelike singularity.}
\end{figure}

\subsection{Region $I$, $(u<0, v<0)$}

This is the flat region with no matter and no gravitational waves.

\subsection{Region $II$, $(u>0, v<0)$}

This is the incoming region from ($+z$), where we define coordinates ($t, z$) as

\begin{equation}
 \sqrt{2}u=t+z, \sqrt{2}v=t-z
\end{equation}
The line element for region II is

\begin{equation}
  ds^2=X^2(u)(2dudv-\cos^2(\frac{au}{\sqrt{1-2k}})dy^2)-\frac{1-\sin(au)}{1+\sin(au)}dx^2,
\end{equation}
where $X(u)=1+\sin(au)$ . It is seen that for a meaningful $g_{yy}$, we must have $au<\frac{\pi}{2}\sqrt{1-2k}$

\subsection{Region $III$, $(u<0, v>0)$}
This is same as region II with $u\leftrightarrow v$ and $a\leftrightarrow b$. It represents our incoming source from ($-z$). Similar to region II, we must have $bv<\frac{\pi}{2}\sqrt{1-2k}$.

\subsection{Region $IV$, $(u>0, v>0)$}

This represents the interaction (or collision) region given by the line element

\begin{equation}
  ds^2=X^2(u)(2dudv-\cos^2(\frac{au-bv}{\sqrt{1-2k}})dy^2)-\frac{1-\sin(au+bv)}{1+\sin(au+bv)}dx^2,
\end{equation}
where $X(u, v)=1+\sin(au+bv)$ and we must have the restriction  $au+bv<\pi\sqrt{1-2k}$, which follows from the conditions of the incoming regions. In the ($t, z$) coordinates, this amounts to  $(a+b)t+(a-b)z<\sqrt{2}\pi\sqrt{1-2k}$.

The Ricci and Weyl components in the Appendix D reveal about the singularities of the solution. The occurrence of Delta delta functions concerns about the singularities on the null cone. When we consider the interaction region alone for $u>0, v>0$, the null singularities are discarded. In the interaction region, the indicative term is the one that is included in all the tetrad scalars, namely the combination $\sim\frac{1}{X^2}$ for Ricci’s and $\frac{1}{X^3}$ for Weyl’s scalars. We investigate the combination $\phi_{22}\sim\frac{1}{X^2}$ in details which describes the energy distribution after the collision. The function $X(u, v)$ involves the $\sin(au+bv)$, which is periodic, however, due the singularities at $X=0$, it amounts to $\sin(au+bv)=-1$, we must avoid this particular hypersurface. For this reason, to stay away from the singularities, we impose the restriction $au+bv<\pi\sqrt{1-2k}$ or equivalently $(a+b)t+(a-b)z<\sqrt{2}\pi\sqrt{1-2k}$. We plot the details of the function $\frac{1}{X^2}$ in both the null $(u, v)$ and the $(t, z)$ coordinates to see the evolution of the Ricci component after the collision.

Our first observation about the Ricci component $\Phi_{22}$ (and $\Phi_{00}$) is that the coefficient of the delta function, namely the function $F(u)=\tan(au)-\frac{1}{\sqrt{1-2k}}\tan(\frac{au}{\sqrt{1-2k}})$ and similarly $G(v)$ with $a\leftrightarrow b$ and $u\leftrightarrow v$, vanish in the incoming regions. This simply means that in the entire allowable domain $au<\frac{\pi}{2}\sqrt{1-2k}$ and $bu<\frac{\pi}{2}\sqrt{1-2k}$  the delta function terms are created and survive on the null boundary. Figure 3.4 and 3.5 display the relevant Ricci components of colliding sources after collision in different coordinate systems.

\begin{figure}[h]
\centering
  \begin{tabular}{@{}cccc@{}}
    \includegraphics[width=0.5\textwidth]{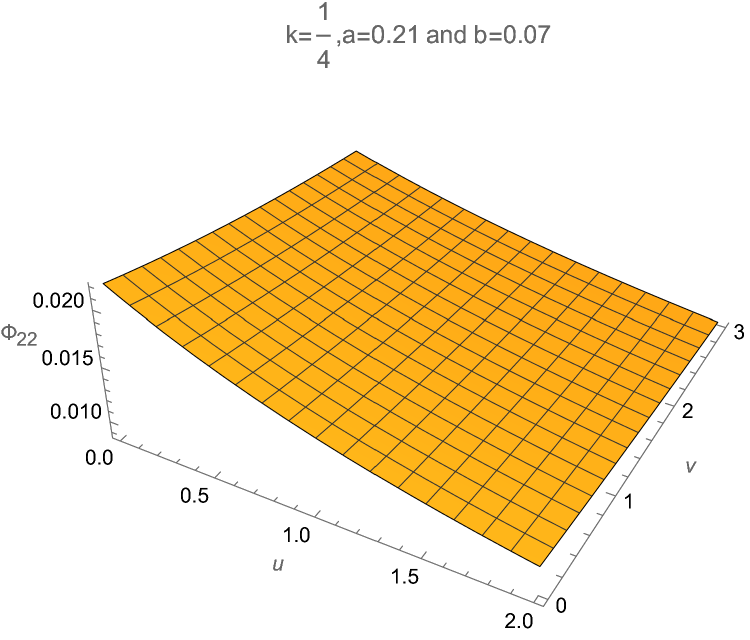} &
    \includegraphics[width=0.5\textwidth]{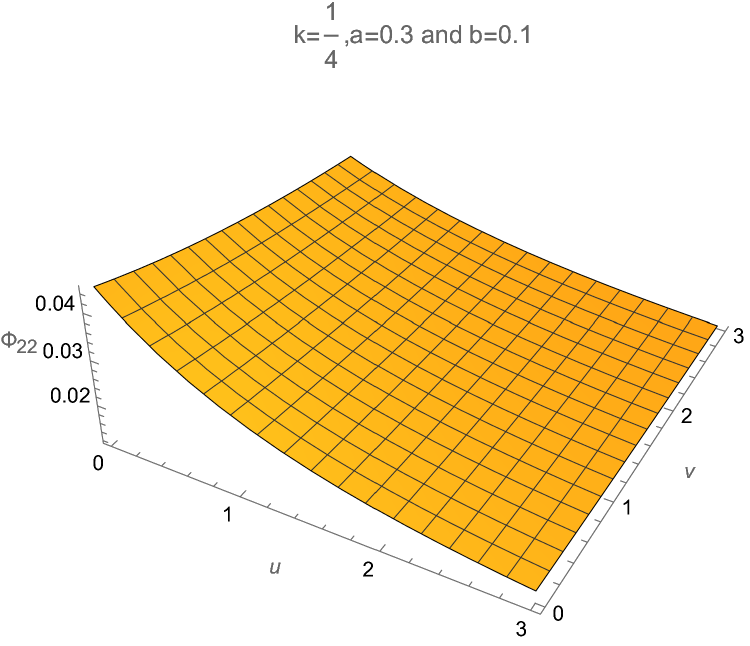} &

                                                             \\
    \multicolumn{2}{c}{\includegraphics[width=0.5\textwidth]{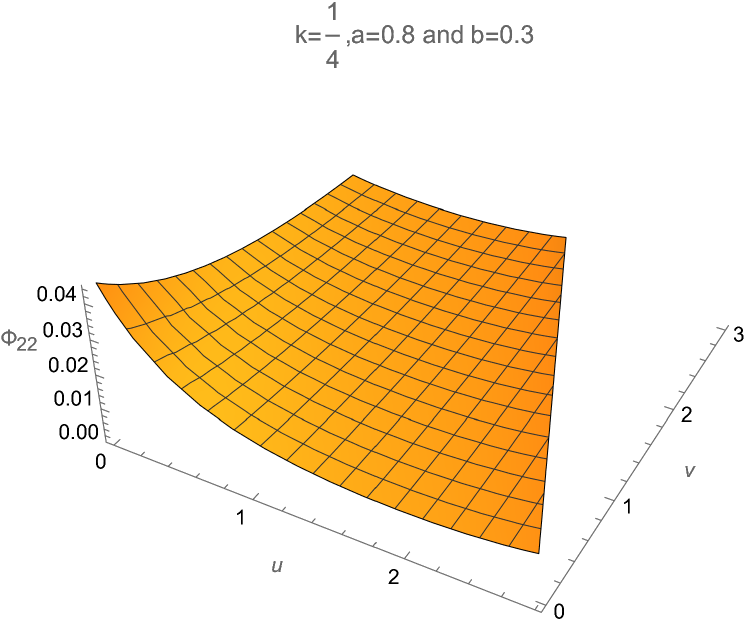}}
  \end{tabular}
  \caption{\centering Post collision plot of the Ricci component $\Phi_{22}=\frac{ka^{2}}{(1-2k)X^{2}}$, for $u>0$, $v>0$ and $k=1/4$, with $X=1+sin(au+bv)$ satisfying the constraint $au+bv<\pi\sqrt{1-2k}$. Note that the plots of  $\Phi_{00}(u,v)$ and $\Phi_{02}(u,v)$ also will follow similar behaviour in which starting from the collision point $u=v=0,$ the Ricci components decrease gradually.}
\end{figure}

\begin{figure}
\centering
\begin{subfigure}{0.45\textwidth}
    \includegraphics[width=\textwidth]{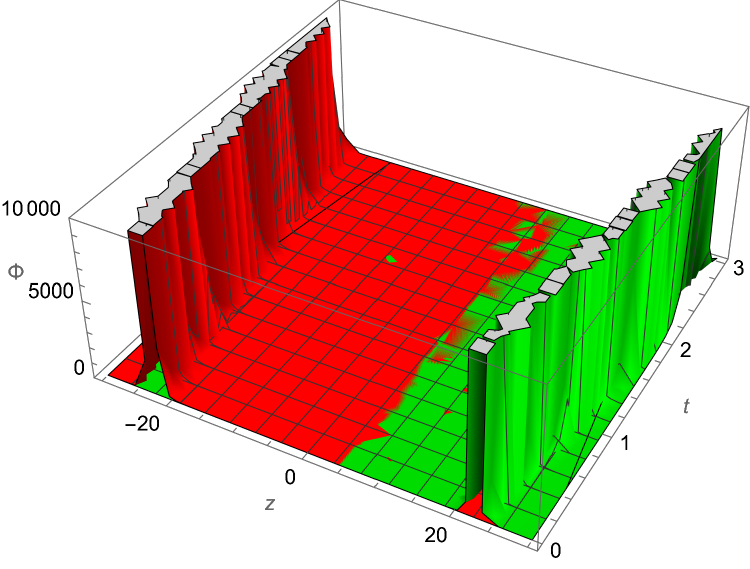}
    \caption{}
    \label{fig:first}
\end{subfigure}
\hfill
\begin{subfigure}{0.45\textwidth}
    \includegraphics[width=\textwidth]{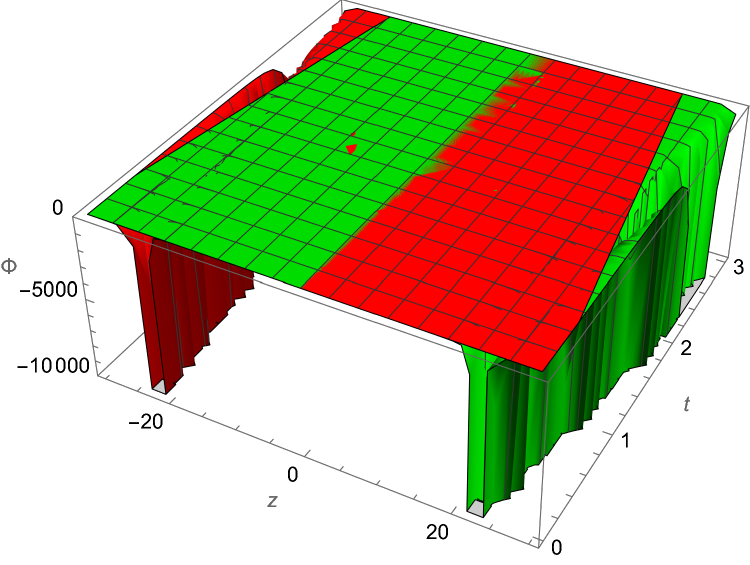}
    \caption{}
    \label{fig:second}
\end{subfigure}

\caption{\centering (a) Plot of the Ricci components $\Phi_{22}(t,z)$ and $\Phi_{00}(t,z)$ in the $(t,z)$ coordinates satisfying the constraint $(a+b)t+(a-b)z<\sqrt{2}\pi\sqrt{1-2k}$. The choice of parameters are $a=0.2$, $b=0.1$ and $k=1/4$. Note that interchanging the parameters $a\leftrightarrow b$ gives $\Phi_{22}\leftrightarrow \Phi_{00}$. (b) Plot of the Ricci component $\Phi_{02}(t,z)$ under same parameters as adapted in part (a) with $a\leftrightarrow b$.}
\label{fig:figures}

\end{figure}

\section{The Physical Interpretation of Our Spacetime}

\subsection{Collision of Pure Gravitational Waves}

In order to understand the present problem, we wish to review briefly the collision of gravitational waves that result locally in the Schwarzchild metric. From region II and III, we have the incoming Weyl components

\begin{equation}
          \Psi_4(u)=a\delta(u)-\frac{3a^2\Theta(u)}{(1+sin(au))^3}
\end{equation}

\begin{equation}
          \Psi_0(v)=b\delta(v)-\frac{3b^2\Theta(v)}{(1+sin(bv))^3}
\end{equation}

respectively, where as before $(a, b)$ are constants. Upon collision we have

\begin{equation}
          \Psi_4(u, v)=\frac{a\delta(u)}{cos(v)(1+sin(bv))^2}-\frac{3a^2\Theta(u)}{(1+sin(au+bv))^3}
\end{equation}

\begin{equation}
          \Psi_0(u, v)=\Psi_4(u\leftrightarrow v, a\leftrightarrow b)
\end{equation}

\begin{equation}
          \Psi_2(u, v)=\frac{ab\theta(u)\theta(v)}{cos(v)(X^3}
\end{equation}

with $X=1+sin(au+bv)$. In the interaction region $u>0, v>0$, the impulsive terms are ignored and what remains satisfy the type-D condition

\begin{equation}
          9\Psi_2^2=\Psi_0\Psi_4
\end{equation}

as it should. Projection of the metric into the $x=const$, $y=const$ plane gives

\begin{equation}
          ds^2=2(1+sin(au+bv))dudv
\end{equation}

which is simple enough to study the Penrose diagram. For $au+bv=\frac{\pi}{2}$, we have $r=2$, which corresponds to the event horizon of the static spacetime. For $au+bv=\pi$, we have $r=1$ and for $au+bv=\frac{3\pi}{2}$, we get $r=0$, the spacelike singularity (Fig. 3.5a,b). In the pure gravitational wave collision problem therefore that the entire interaction region takes place in between $0<r<2$. For more details of this particular case, we refer \cite{griffiths2016colliding}.

\subsection{Collision of Null Matter Source}

By introducing an incoming null source with the parameter $k\not = 0$, from the theory of bumblebee gravity many features encountered in previous section needs revision. First of all, in the incoming region II and III, in order to define a physical source, such as $\Phi_{22}>0$, and $\Phi_{00}>0$, we must choose $k<\frac{1}{2}$. This automatically has restrictions on the acceptable range of the null coordinates u and v. By the local isometry the metric (3.23) obtained from the static one and the NP curvature/Ricci components (see Appendix D) suggest that we must choose $au<\frac{\pi}{2}\sqrt{1-2k}$ and $bv<\frac{\pi}{2}\sqrt{1-2k}$. As a result, we end up with the limitation

\begin{equation}
          au+bv<\frac{\pi}{2}\sqrt{1-2k}.
\end{equation}
Now, projecting the metric to the $(u, v)$ plane gives us the same metric form as in (3.40). Unlike the case of colliding pure gravitational waves isometric to the Schwarzchild geometry the interaction region now is restricted by the condition $1<1+sin(\pi\sqrt{1-2k})$, which is applicable only for $u>0$, $v>0$ and $k\neq0$. The event horizon of the vacuum $r=2$, is not accessible in the present case. Given the condition (3.41), the data for the ejected jets from the Ricci tensor components are bound to stay inside the horizon and in particular in the shaded region of figure 3.6 (c). It should be reminded that already the jets take place in the vicinity of black holes and any horizon to form becomes time dependent, shrinking in time. In the absence of horizon or in the presence of naked singularities, the observability of the forming jets becomes more clear.

\begin{figure} 
    \centering
    \subfloat[\centering ]{{\includegraphics[scale=0.3]{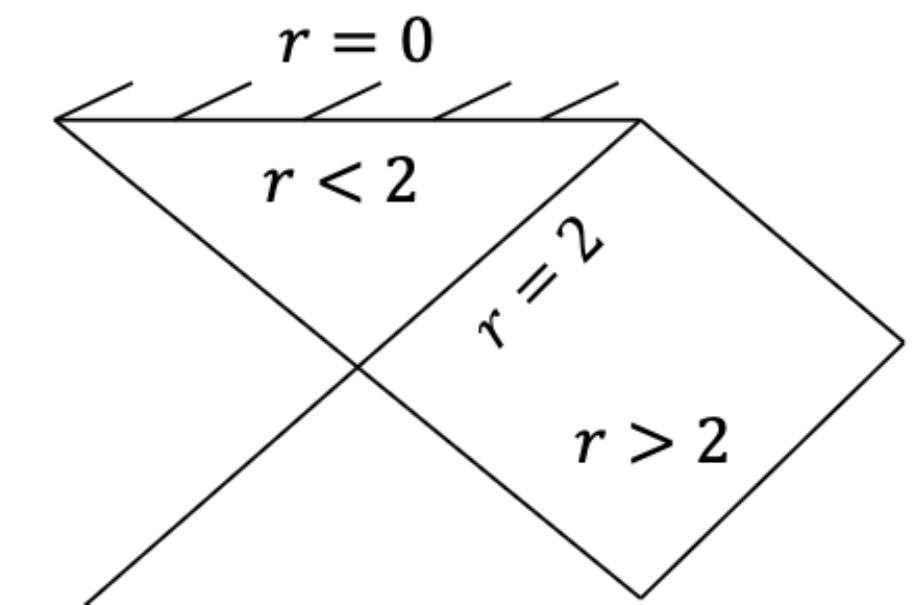} }}%
    \qquad
    \subfloat[\centering ]{{\includegraphics[scale=0.3]{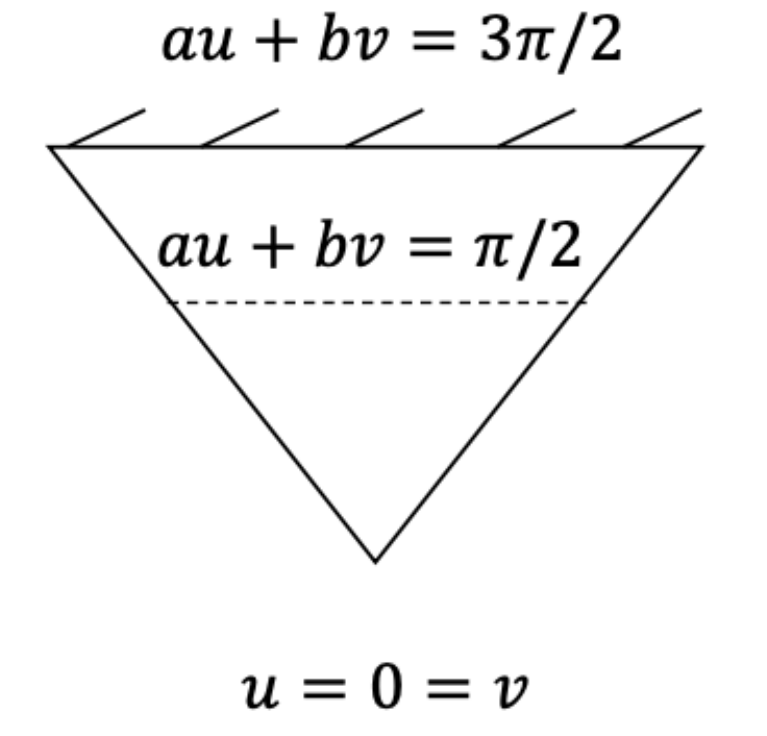} }}%
    \qquad
    \subfloat[\centering  ]{{\includegraphics[scale=0.2]{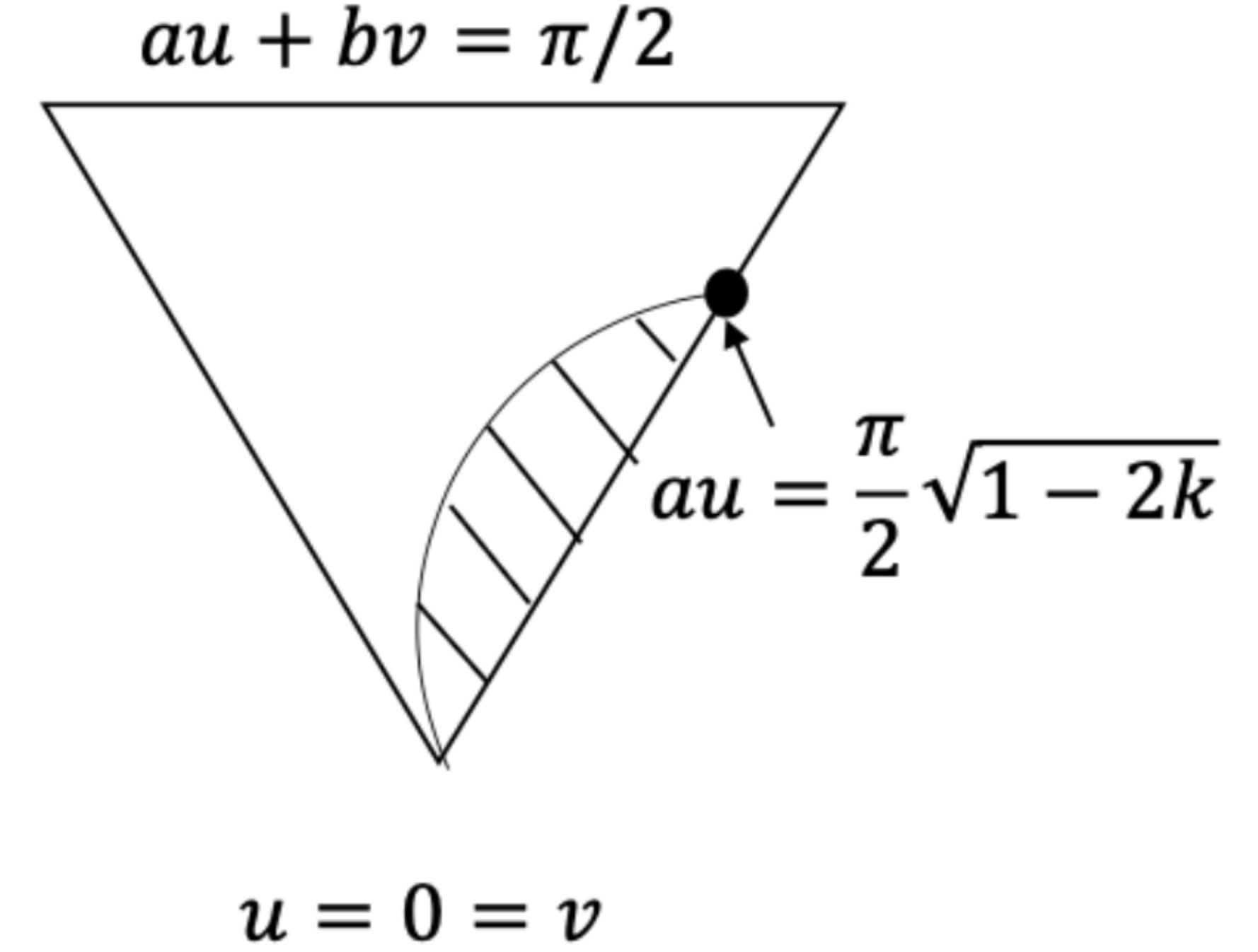} }}%
    
    \caption{\centering (a) Part of the Penrose diagram representing the static spacetime for $0\leq<\infty$ with $m=1$. The role of the parameter $k$ is to add a constant to the tortoise coordinate in the Schwarzchild spacetime. Our region of collision occurs in the region denoted by $0<r<2$. (b) Penrose diagram for the interaction region of colliding gravitational waves isometric to the Schwarzchild spacetime from the instant of collision $u=v=0$, $au+bv=\frac{\pi}{2}$ is a horizon where $au+bv=\frac{3\pi}{2}$ corresponds to the spacelike singularity of static spacetime $r=0$. (c) Penrose diagram for colliding null matter represented by the line element (12) in the text. The singularity is reached earlier at the indicated heavy dot point and its symmetric point with $u \leftrightarrow v$, $a \leftrightarrow b$. The reason may be attributed to the stronger focusing by the null matter relative to the pure gravitational wave problem. Geodesic particles and jets emerge all in the shaded region and its symmetric region. }%
    \label{fig1}%
\end{figure}

\section{Summary}
In this study, we investigate whether astrophysical jets observed in the context of general relativity can be described using a restricted approach. Given that jets originate from strongly attractive sources like black holes, we focus on the collision of oppositely moving sources from accretion disks. This collision process requires symmetry to achieve an exact solution, as the attracted sources approach nearly the speed of light, the null matter. While colliding waves are a well-established topic in general relativity \cite{griffiths2016colliding}, our aim is to understand jet formation through this phenomenon. However, reaching the speed of light introduces undesired singularities, which arise before the expected gravitational wave singularities. And a redistribution of the source leads to the emergence of impulsive null shells characterized by Dirac delta functions. The subject of singularities and their possible removal constitutes by itself a separate title \cite{griffiths2016colliding,clarke1989global,newman1962approach,ellis1977singular} which is not addressed in this study

Our source model involves a global monopole coupled with a Schwarzchild source or equivalently, a cloud of strings or a static source in bumblebee gravity. This choice is based on simplicity, as it represents the simplest extension of the Schwarzchild source; that is, the existence of k, or we can say the Lorentz symmetry breaking constant l, makes the discussion and calculation accessible. But when k=0, jet-like formations disappear, reducing the problem to colliding gravitational waves, which is isometric to the Schwarzschild geometry. Besides, one should notice that we utilize static line elements to capture the dynamics within the event horizon, where spacetime transitions from static to dynamic. One can see the implication from the coordinate transformation (3.18-21).

Finally, our method can be applied to any Schwarzschild-like metric. Although our model simplifies certain physical factors, such as additional polarizations or electromagnetic fields, it offers insights into the formation of astrophysical jets.

\chapter{The null-source collision and confinement in the 
 Lovelock gravity of the third order}\label{ch:preliminary}

 In this paper, we extend the idea of the 4-dimensional Gauss-Bonnet (GB) theory to the third-order Lovelock
gravity (TOLG) and obtain a general class of
static spherically symmetric solutions. These solutions encompass the Schwarzschild-de Sitter limit and a broad class of other possibilities. However, these solutions do not accommodate flat rotation curves. Remarkably, the simplest member of the solution class corresponds to a specific instance of the Mannheim and Kazanas \cite{mannheim1989exact} static solution, previously derived within the framework of General Relativity under different circumstances. We then study the confinement of radial geodesics within this simplest TOLG metric, with a particular focus on the pure Lovelock gravity restricted to the third order. While this class of metrics hosts an infinite class of static solutions, it is notable that not every element with fractional power corresponds to a well-known physical potential. Nonetheless, each power represents a distinct source configuration, enriching the physical interpretation of the solutions.

Subsequently, we transform the static metric into double-null coordinates to study the collision of physical sources. We notice that two incoming conformally flat (CF) sources collide and then yield a non-CF interaction region. This behavior is in contrast to the collision of null electromagnetic waves.

\section{Lovelock Gravity: A Brief Introduction}

The Lovelock gravity is followed as \cite{lovelock1971einstein}%

\begin{equation}
S=\frac{1}{2\kappa ^{2}}\int d^{d}x\sqrt{-g}\sum_{k=0}^{\left[ \frac{d-1}{2}%
\right] }\alpha _{k}\emph{L}_{k},  \label{1}
\end{equation}%

which is an action of $d$ dimensions, where $\alpha _{k}$ are arbitrarily real constants, $\left[ \frac{d-1}{2}%
\right] $ is the integral part of $\frac{d-1}{2}$ and

\begin{equation}
\emph{L}_{k}=\frac{1}{2^{k}}\delta _{\mu _{1}\nu _{1}...\mu _{i}\nu
_{i}}^{\alpha _{1}\beta _{1}...\alpha _{i}\beta
_{i}}\prod\limits_{i=1}^{k}R_{\text{ \ \ \ \ \ \ \ }\alpha _{i}\beta
_{i}}^{\mu _{i}\nu _{i}}  \label{2}
\end{equation}%

are the Euler densities.

In this chpater we shall focus on the third order
Lovelock Lagrangian, $k=3$, that is,

\begin{multline}
\emph{L}_{3}=R^{3}-12RR^{\mu \nu }R_{\mu \nu
}+16R^{\mu \nu
}R_{\nu \sigma }R_{\hspace{0.15cm}\mu }^{\sigma }+2R^{\kappa \lambda \rho \sigma }R_{\rho \sigma \mu \nu }R_{%
\hspace{0.3cm}\kappa \lambda }^{\mu \nu }\\
+8R_{\hspace{0.3cm}\kappa \lambda
}^{\mu \nu }R_{\hspace{0.3cm}\nu \rho }^{\kappa \sigma }R_{\hspace{0.3cm}\mu
\sigma }^{\lambda \rho }+24R^{\kappa \lambda \mu \nu }R_{\mu \nu \lambda\rho }R_{\hspace{0.15cm}\kappa }^{\rho }+3RR^{\kappa \lambda \mu \nu}R_{\kappa \lambda \mu \nu }\\
+24R^{\kappa \lambda \mu \nu }R_{\mu \kappa }R_{\nu \lambda },\label{4}  \notag
\end{multline}%

with the third order Lovelock parameter, $\alpha _{3}$. Then considering $d$%
-dimensional spherically symmetric static spacetime with line element%

\begin{equation}
ds^{2}=-f\left( r\right) dt^{2}+\frac{dr^{2}}{f\left( r\right) }%
+r^{2}d\Omega _{d-2}^{2},  \label{5}
\end{equation}%
the Einstein-Lovelock's field equation reduces to a $k$-order ordinary
equation given by%

\begin{equation}
\sum_{k=0}^{\left[ \frac{d-1}{2}\right] }\tilde{\alpha}_{k}\psi ^{k}=\frac{%
16\pi M}{\left( d-2\right) \Sigma _{d-2}r^{d-1}},  \label{6}
\end{equation}%

where $f=1-r^{2}\psi $ and the integration constant $M$ represents the mass of
the asymptotically flat black hole or non-black hole solution, and also

\begin{equation}
\Sigma _{d-2}=\frac{2\pi ^{\frac{d-1}{2}}}{\Gamma \left( \frac{d-1}{2}%
\right) }.  \label{7}
\end{equation}%

$\bar{\alpha}%
_{0}=\frac{\alpha _{0}}{\left( d-1\right) \left( d-2\right) }$, $\bar{%
\alpha}_{1}=1$ and for $k\geq 2$

\begin{equation}
\bar{\alpha}_{k}=\prod\limits_{i=3}^{2k}\left( d-i\right) \alpha _{k}.
\label{8}
\end{equation}%

In $d=4,$ for $k\geq 2$, $\bar{\alpha}_{k}=0$ and Eq. (\ref{6})
becomes%

\begin{equation}
\frac{\alpha _{0}}{6}+\psi =\frac{2M}{r^{3}},  \label{9}
\end{equation}%

which is the Schwarzschild-de Sitter solution \cite{mazharimousavi2023confinement}, where %

\begin{equation}
f=1+\frac{\alpha _{0}}{6}r^{2}-\frac{2M}{r}.  \label{10}
\end{equation}%

Next, we introduce the new Lovelock parameters $\beta _{k}=$ $\frac{\alpha
_{k}}{\prod\limits_{i=3}^{2k}\left( d-i\right) }$ in which $\alpha _{k}$ are
real constants. This, however implies that the action (\ref{1}) becomes as%

\begin{equation}
S=\frac{1}{2\kappa ^{2}}\int d^{d}x\sqrt{-g}\sum_{k=0}^{m}\beta _{k}\emph{L}_{k},  \label{11}
\end{equation}%

where $m$ is any positive integer number. Therefore, the Einstein-Lovelock field
equation becomes%

\begin{equation}
\sum_{k=0}^{m}\alpha _{k}\psi ^{k}=\frac{16\pi M}{\left( d-2\right) \Sigma
_{d-2}r^{d-1}}.  \label{12}
\end{equation}%

In the limit of $d\rightarrow 4$, it takes the form \cite{konoplya20204d}%

\begin{equation}
\sum_{k=0}^{m}\alpha _{k}\psi ^{k}=\frac{2M}{r^{3}}.  \label{13}
\end{equation}%
This formula is what we are interested.

\section{The Pure Third-order Lovelock
gravity (TOLG)}
 
We may set only $\alpha _{k=p}=\alpha$ to be nonzero in Eq. (\ref{13}), which implies%

\begin{equation}
\alpha \psi ^{p}=\frac{2M}{r^{3}}.  \label{19}
\end{equation}%

The solution for the metric function reads%

\begin{equation}
f\left( r\right) =1-r^{\frac{2p-3}{p}}\left( \frac{2M}{\alpha }\right) ^{%
\frac{1}{p}}.  \label{20}
\end{equation}%

One should impose $\frac{2M}{\alpha }>0$ for an even integer value of $p$. However for an odd value for $p,$ it can be positive or negative. Since we
are interested in confinement \cite{mazharimousavi2023confinement} we set $p=3$ and choose $\frac{2M}{\alpha }<0$
which bring us%

\begin{equation}
\text{i) }f\left( r\right) =1+2ar\text{ for }\frac{2M}{\alpha }<0,  \label{21}
\end{equation}%
and%

\begin{equation}
\text{ii) }f\left( r\right) =1-2ar\text{ for }\frac{2M}{\alpha }>0,
\label{22}
\end{equation}%
in which $2a=\left\vert \frac{2M}{\alpha }\right\vert ^{\frac{1}{3}}.$ The metric reads%

\begin{equation}
ds^{2}=-\left( 1\pm 2ar\right) dt^{2}+\frac{dr^{2}}{\left( 1\pm 2ar\right) }%
+r^{2}d\Omega ^{2},  \label{23}
\end{equation}%

which is non asymptotically flat solution and $a$ represents a source, as we
shall show below. This metric is a special form of the Mannheim-Kazaras \cite{mannheim1989exact} spacetime.

\section{Geodesics Confinement}
We then consider the radial geodesics for $\theta =$constant, $\varphi =$constant, so
that the Lagrangian of the metric \cite{rindler2006relativity} reduces to

\begin{equation}
\mathcal{L}=-\frac{1}{2}f\dot{t}^{2}+\frac{1}{2f}\dot{r}^{2},  \label{24}
\end{equation}%

where $f=1-2ar$ and a 'dot' denotes derivative with respect to the proper
time $\tau $. We note that the results can easily be modified with the
change of sign in the source parameter, $a\rightarrow -a.$ The
Euler-Lagrange equations imply

\begin{equation}
f\frac{dt}{d\tau }=E=constant  \label{25}
\end{equation}%
and%
\begin{equation}
\dot{r}^{2}=E-f.  \label{26}
\end{equation}%
So that the second derivatives read

\begin{equation}
\frac{d^{2}r}{d\tau ^{2}}=a  \label{27}
\end{equation}%
and
\begin{equation}
\frac{d^{2}r}{dt^{2}}=-2a\left( 1-2ar\right) \left[ 1-\frac{3}{2E^{2}}\left(
1-2ar\right) \right] .  \label{28}
\end{equation}%
For $r\rightarrow 0,$ we have

\begin{equation}
\frac{d^{2}r}{dt^{2}}\sim \frac{3a}{E^{2}}-2a  \label{29}
\end{equation}%
and also for $r\rightarrow \infty $%
\begin{equation}
\frac{d^{2}r}{dt^{2}}\sim 12\frac{a^{3}}{E^{2}}r^{2}.  \label{30}
\end{equation}%

For $r\rightarrow 0,$ we have ($a>0$)%

\begin{equation}
\frac{d^{2}r}{dt^{2}}=\left\{
\begin{array}{cc}
\text{attractive force,} & \text{for }E^{2}>\frac{3}{2} \\
\text{repulsive force,} & \text{for }E^{2}<\frac{3}{2}%
\end{array}%
\right.  \label{31}
\end{equation}%
and for $r\rightarrow \infty ,$ we have only a repulsive force. For $%
a\rightarrow -a$, these forces all reverse sign, i.e., attractive becomes
repulsive and vice versa. The general integral integral for $r\left(
t\right) $ is optional from (\ref{31}) as

\begin{equation}
r\left( t\right) =\frac{1}{2a}\left( 1-\frac{E^{2}}{\cosh ^{2}at}\right),
\label{32}
\end{equation}%
which implies that for $0<t<\infty $ the particle is confined by%

\begin{equation}
-E^{2}+1<2ar<1.  \label{33}
\end{equation}%
By letting $a\rightarrow -\left\vert a\right\vert ,$ in the TOLG, we see
that the radial geodesics will give an upper bound

\begin{equation}
2\left\vert a\right\vert r=-1+E^{2}  \label{34}
\end{equation}%
and this is confined for a finite energy.

\section{No Flat Rotation Curves}
Can our choice correspond to a cosmological phenomenon? 
From Eq. (\ref{20}) with the choice $2a=\left( \frac{2M}{\alpha }\right) ^{1/P}$ we
have the metric function $g_{tt}$ of the form.

\begin{equation}
g_{tt}=1-2ar^{2-\frac{3}{P}}.  \label{35}
\end{equation}%

The  Newtonian potential, $\Phi _{N}$, is
defined by $g_{tt}=1+2\Phi _{N}.$ For example, $P=1,$ we have $\Phi
_{N}=-\frac{a}{r},$ which implies that the parameter $a$ is the mass. And the Newtonian force is defined by

\begin{equation}
F_{N}=-\frac{d\Phi _{N}}{dr}=-\frac{a}{r^{2}}.  \label{36}
\end{equation}%
We tabulate the Newtonian forces for different $P-$values as follows%

\begin{equation}
F_{N}=\left\{
\begin{array}{cc}
\frac{constant}{r^{2}}, & \text{for }P=1,\text{ Newton} \\
\frac{constant}{\sqrt{r}}, & \text{for }P=2,\text{ GB} \\
constant & \text{for }P=3,\text{ pure Lovelock} \\
... & ... \\
\left( constant\right) \sqrt{r} & \text{for }P=6,\text{ etc.}%
\end{array}%
\right. .  \label{37}
\end{equation}%
For the foregoing list of forces it can be shown that the flat
rotation curve (FRC) condition is not admitted. We consider
the speed of circular velocity $v_{c}$ to see this, which must satisfy

\begin{equation}
v_{c}^{2}=r\left\vert \frac{d\Phi _{N}}{dr}\right\vert =constant\text{.}
\label{38}
\end{equation}%
We show the forgoing cases in the following results

\begin{equation}
v_{c}\sim \left\{
\begin{array}{cc}
\frac{1}{r}, & P=1 \\
r^{1/4}, & P=2 \\
r^{1/2}, & P=3 \\
r^{5/8}, & P=4 \\
...$so forth and so on$.
\end{array}%
\right.  \label{39}
\end{equation}%
In order that the FRC will be satisfied the constant $a$ (or mass)
appearing in $\Phi _{N}$ must be a function of $r,$ which implies for the
Newtonian case that the enclosed mass must be a linear function of $r,$
which is not our case. In this sense the above examples fail to support a FRC.

\section{ Colliding Fields in the Particular TOLG}

We shall impose the colliding tricks in this section. One may consider the TOLG metric in the form

\begin{equation}
ds^{2}=-fdt^{2}+\frac{dr^{2}}{f}+r^{2}\left( d\theta ^{2}+\sin ^{2}\theta
d\varphi ^{2}\right),   \label{40}
\end{equation}%
where $f=1-2ar$ with $a>0.$ To see the contend of this static metric we
define the proper null-tetrad basis of Newman-Penrose (NP)%

\begin{equation}
\ell _{\mu }=\left( 1,-\frac{1}{f},0,0\right),   \label{41}
\end{equation}%

\begin{equation}
n_{\mu }=\frac{1}{2}\left( f,1,0,0\right),   \label{42}
\end{equation}%

\begin{equation}
m_{\mu }=-\frac{r}{\sqrt{2}}\left( 0,0,1,i\sin \theta \right),   \label{43}
\end{equation}%

\begin{equation}
\bar{m}_{\mu }=-\frac{r}{\sqrt{2}}\left( 0,0,1,-i\sin \theta \right),
\label{44}
\end{equation}%
with the convention%

\begin{equation}
g_{\mu r}=-\ell _{\mu }n_{r}-\ell _{r}n_{\mu }+m_{\mu }\bar{m}_{r}+m_{r}\bar{%
m}_{\mu }.  \label{45}
\end{equation}%
This brings the non-zero NP quantities%

\begin{equation}
\psi _{2}=-\frac{a}{r}  \label{47}
\end{equation}%

\begin{equation}
\Phi _{11}=\frac{a}{2r}  \label{46}
\end{equation}%

\begin{equation}
\Lambda =\frac{R}{24}=\frac{a}{r}.  \label{48}
\end{equation}%

It is seen that the parameter $a>0,$ creates both source and conformal
curvature. Now, we transform our static metric (\ref{40}) into the double
null-coordinates ($u,v$), as follows

\begin{equation}
r=\frac{1}{F}  \label{49}
\end{equation}%
\begin{equation}
t=x  \label{51}
\end{equation}%
\begin{equation}
\varphi =y  \label{52}
\end{equation}%
\begin{equation}
\theta =u-v+\frac{\pi }{2}  \label{50}
\end{equation}%

where $F=a\left( 1-\sin \left( u+v\right) \right) .$ The resulting metric, after a scaling on $x$ and $y$ will be

\begin{equation}
ds^{2}=\frac{1}{F^{2}}\left[ -2dvdu+a^{2}\cos ^{2}\left( u+v\right)
dx^{2}+\cos ^{2}\left( u-v\right) dy^{2}\right]   \label{53}
\end{equation}%
which is still identical with the conformally non-flat line element (\ref{40}). The next step we shall formulate the static metric as resulting from
appropriate colliding fields. We make the Penrose substitution as we did in previous chapter%

\begin{equation}
u\rightarrow u\theta \left( u\right)   \label{54}
\end{equation}%
\begin{equation}
v\rightarrow v\theta \left( v\right).   \label{55}
\end{equation}%

We consider now the spacetime divided into four regions \cite{tahamtan2016colliding}\cite{bell1974interacting}\cite{halilsoy1988colliding}\cite{wu1982scalar}\cite{mazharimousavi2020corrigendum}, as (see figure 4.1)%

\begin{equation}
\left.
\begin{array}{c}
\text{Flat Region I, (}u<0,v<0\text{)} \\
\text{Incoming Region II, (}u>0,v<0\text{)} \\
\text{Incoming Region III, (}u<0,v>0\text{)} \\
\text{Interaction Region IV, (}u>0,v>0\text{)}%
\end{array}%
\right.   \label{57}
\end{equation}

The incoming region II ($v<0$) is specified by the line element%

\begin{equation}
ds^{2}=\frac{1}{F^{2}\left( u\right) }\left[ -2dvdu+\cos ^{2}\left( u\theta
\left( u\right) \right) \left( a^{2}dx^{2}+dy^{2}\right) \right]   \label{58}
\end{equation}%
which is seen to be CF upon redefinition of the null coordinate $u$. The
incoming region III ($u<0$) is similar to this with $u\leftrightarrow v$.
The collision of these two CF metrics gives rise to the interaction metric%

\begin{equation}
ds^{2}=\frac{1}{F^{2}\left( u,v\right) }\left[ -2dvdu+a^{2}\cos ^{2}\left(
u\theta \left( u\right) +v\theta \left( v\right) \right) dx^{2}+a^{2}\cos
^{2}\left( u\theta \left( u\right) -v\theta \left( v\right) \right) dy^{2}%
\right]   \label{59}
\end{equation}%
in which $F\left( u,v\right) =a\left( 1-\sin \left( u\theta \left( u\right)
+v\theta \left( v\right) \right) \right) .$ The choice of NP null-tetrad%

\begin{equation}
\ell _{\mu }=\frac{1}{F}\delta _{\mu }^{u}  \label{60}
\end{equation}%

\begin{equation}
n_{\mu }=\frac{1}{F}\delta _{\mu }^{v}  \label{61}
\end{equation}%

\begin{equation}
m_{\mu }=\frac{-1}{\sqrt{2}F}\left[ a\cos \left( u\theta \left( u\right)
+v\theta \left( v\right) \right) \delta _{\mu }^{x}+i\cos \left( u\theta
\left( u\right) -v\theta \left( v\right) \right) \delta _{\mu }^{y}\right]
\label{62}
\end{equation}%

we obtain the following NP quantities in the interaction region, including
the null boundaries%

\begin{equation}
\Phi _{00}=aF\left( u,v\right) \left( \theta \left( v\right) -\delta \left(
v\right) \cos \left( u\theta \left( u\right) \right) \right)   \label{66}
\end{equation}%

\begin{equation}
\Phi _{22}=aF\left( u,v\right) \left( \theta \left( u\right) -\delta \left(
u\right) \cos \left( v\theta \left( v\right) \right) \right)   \label{65}
\end{equation}%

\begin{equation}
\Phi _{02}=\Phi _{20}=-a\theta \left( v\right) \theta \left( u\right)
F\left( u,v\right)   \label{64}
\end{equation}%

\begin{equation}
\Phi _{11}=\frac{1}{8}\theta \left( v\right) \theta \left( u\right)
F^{2}\left( u,v\right) \left( \frac{1}{\cos ^{2}\left( u+v\right) }-\frac{1}{%
\cos ^{2}\left( u-v\right) }\right)   \label{67}
\end{equation}%

\begin{equation}
  \begin{array}{cc}

-3\Lambda =-\frac{R}{8}=aF( u,v) \theta ( v) \theta
( u) [ \cos ( u+v) -\sin ( u+v)  

\\
-3 -\frac{F}{8a( \tan ^{2}( u+v) -\tan ^{2}( u-v)}
]  

  \end{array}
\end{equation}%

\begin{equation}
\Psi _{4}=\delta \left( u\right) F^{2}\left( u,v\right) \tan \left( v\theta
\left( v\right) \right)   \label{69}
\end{equation}%

\begin{equation}
\Psi _{0}=\delta \left( v\right) F^{2}\left( u,v\right) \tan \left( u\theta
\left( u\right) \right)   \label{70}
\end{equation}%

\begin{equation}
\begin{array}{cc}
12\Psi _{2}=-\theta ( v) \theta ( u) F^{2}(
u,v) [ \tan ^{2}( u+v) -\tan ^{2}( u-v) \\
-\frac{4a}{F( u,v) }( \cos ( u+v) -\sin (
u+v) ) ]   \label{71}
\end{array}
\end{equation}
where $\delta \left( u\right)$ and $ \delta \left( v\right) $ is again the Dirac delta
function. It cab be seen that $u+v=\frac{\pi }{2}$ is a singular hypersurface.
The singularity surface $u-v=\frac{\pi }{2}$ lies outside the null core so
it is not of physical interest. Interestingly, there are null singularities
located at ($u=0,v=\frac{\pi }{2}$) and ($v=0,u=\frac{\pi }{2}$). We take $u>0,v>0$ and ignore the delta functions inside the region IV. This shows that the interaction region is not CF. Since $\Psi _{2}\neq 0$ is locally isometric with what we have obtained for
the TOLG, which is equivalent to the Mainnheim's metric \cite{mannheim1989exact}. To sum up,
we have found the foregoing static metric to derive locally from
collision of two appropriate null sources without gravitational waves. It
should be added, collision of two incoming sources here doesn't result in a CF outcome, compared to the more familiar problem of colliding
null electromagnetic fields. In the
electromagnetic field problem on the other hand collision of two CF, null
fields resulted in a non-null but still CF spacetime. Collision of our
null fields therefore is in marked distinction with the colliding electromagnetic
problem.

\begin{figure}
  \centering
  \includegraphics[width=0.8\textwidth]{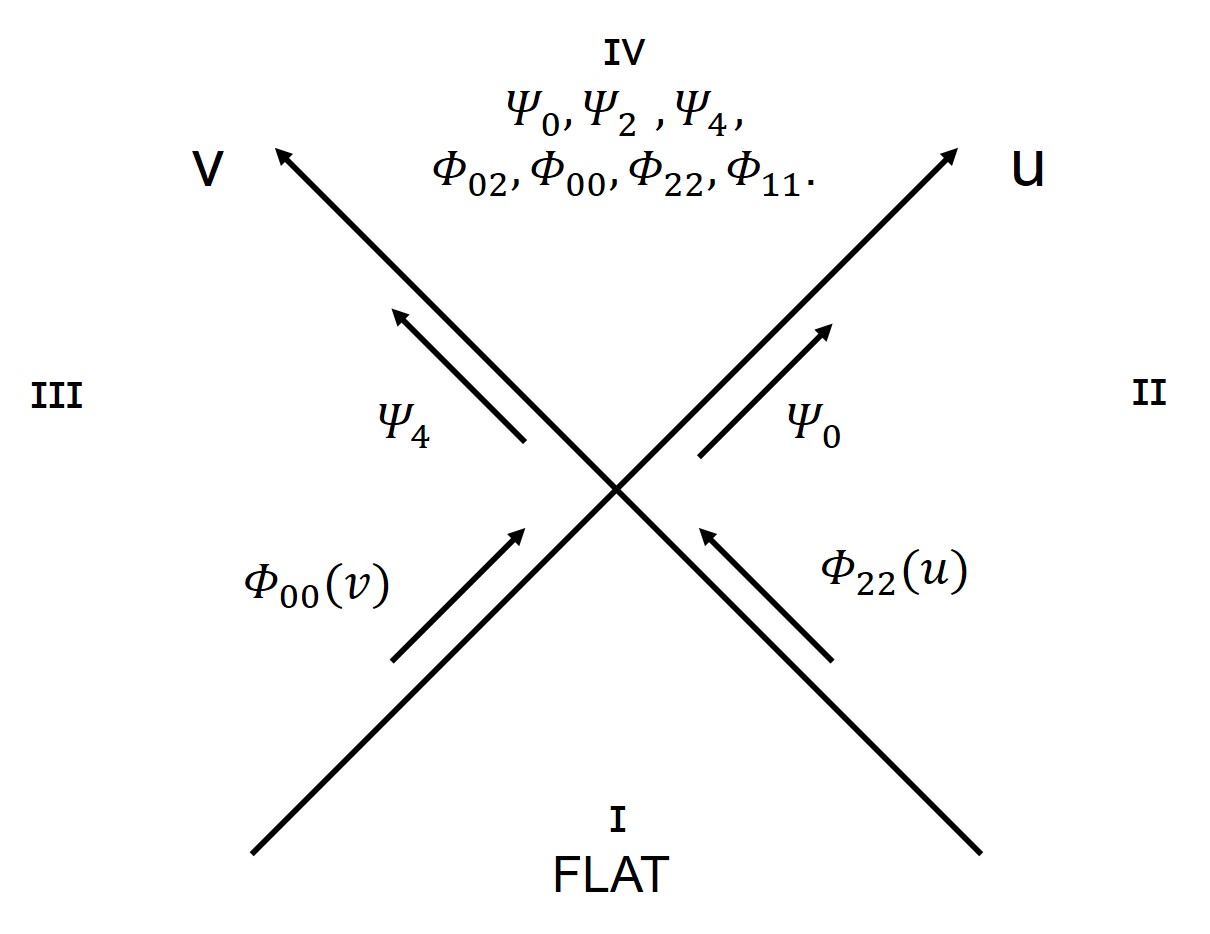}
  \caption{\centering The spacetime diagram for colliding null sources. We have shown all nonzero quantities after the collision. Prior to the collision, there is no conformal curvature. The incoming region II ($v<0$) is specified by the metric
$
ds^{2}=\frac{1}{F^{2}\left( u\right) }\left[ -2dudv+\cos ^{2}\left( u\theta
\left( u\right) \right) \left( a^{2}dx^{2}+dy^{2}\right) \right]
$
which CF upon redefinition of the null coordinate $u$. The incoming region III ($u<0$) is same to this with $u\leftrightarrow v$. The collision of these two CF metrics gives rise to the interaction metric
$
ds^{2}=\frac{1}{F^{2}\left( u,v\right) }\left[ -2dudv+a^{2}\cos ^{2}\left(
u\theta \left( u\right) +v\theta \left( v\right) \right) dx^{2}+a^{2}\cos
^{2}\left( u\theta \left( u\right) -v\theta \left( v\right) \right) dy^{2},%
\right]
$
where $F\left( u,v\right) =a\left( 1-\sin \left( u\theta \left( u\right)
+v\theta \left( v\right) \right) \right) .$
 }\label{f1}
\end{figure}

\section{Summary}

In line with recent developments in the 4D Gauss-Bonnet (GB) theory, we extend this concept to Lovelock gravity, particularly focusing on the pure third-order variant. We introduce a class of solutions characterized by the parameter $0 < p < \infty$. We focus on the specific case where $p = 3$. This choice results in a straightforward static form that aligns with the particular case of the Mannheim-Kazanas metric. By expanding in powers of $1/r$ with the choice $\alpha_0 = 0$, we observe the potential lacking dipole and quadrupole terms, contrasting with the standard general relativity.

Within this static metric framework, we initially examine pure radial geodesics and their confinement. Remarkably, we find that every particle with finite energy becomes confined within the metric time interval extending from negative to positive infinity. Subsequently, we explore the flat rotation curves within the class of solutions derived in TOLG. Despite investigating certain p-values, we note the absence of curves yielding a constant $v^2$ term.

As a third application of our TOLG solution, we investigate the collision of null sources, corresponding to boosted static sources. An interesting outcome emerges as the collision of conformally flat (CF) sources generates a non-CF spacetime, contrasting with the scenario observed in the collision of electromagnetic waves, where CF incoming waves result in a CF interaction region. Additionally, we note that both cases exhibit impulsive Weyl curvatures along the null boundaries after the collision.

\chapter{Conclusion and outlook}\label{ch:preliminary}

After we introduce the basic mathematical techniques 
in chapter 1, we deal with 3 topics, the CGHS model in 1+1-dimensional spacetime, the toy model of astrophysical jets formation in the Schwarzchild-like black holes, and collisions and confinement in the third order of Lovelock gravity. Here are the features and prospects.

We collide the ghost fields in the CGHS background and realize that wormholes form and disassemble when ghost field disappear. This is analogous to the Feynman diagram but in a cosmological scale. We also find that the wormholes may show up by putting electromagnetic fields inside the Lagrangian. This field can not be replaced by Penrose substitution. So we did not include it in our argument. We shall discuss it in the future.

Next, the non-zero expectation value of bumblebee fields arise when Lorentz symmetry breaks. The vacuum in energy-momentum tensor now should include the potential vacuum, therefore the vacuum solution will shift to become Schwarzchild-like black holes with a factor $l$. This factor provides the advantage of discussing the collision of null sources. A redistribution of the source after the collision leads to the emergence of impulsive null shells. And the type-D condition is satisfied in the interacting region. When $l=0$, jet-like formations disappear, reducing the problem to colliding gravitational waves, which is isometric to the Schwarzschild geometry. Furthermore, our method can be applied to any Schwarzschild-like metric. We wish our model can include some certain physical factors, such as additional polarizations or electromagnetic fields.

Finally, we discuss the 4-dimensional third-order Lovelock
gravity. We find that every particle with finite energy becomes
confined within the metric time interval $-\infty\rightarrow\infty$, but no flat rotation curves are admitted. Then making collision under the background of this metric, interestingly, impulsive Weyl curvatures emerge along the null boundaries after the
collision. By adjusting the parameters in the third order Lovelock gravity, We hope that more geometries with physical meaning can be found.

\bibliographystyle{IEEEtran}
\bibliography{references}


\begin{appendices}

\appndxsec{How to Derive Dual Forms}{\label{app:importantnotes}}

In the orthonormal frame, we define the metric as $ds^2=-(\omega^0)^2+(\omega^1)^2+(\omega^2)^2+(\omega^3)^2$ and the c-number 1 as 0-form. The dual of 1 is $*1=\omega^0\wedge\omega^0\wedge\omega^0\wedge\omega^0$, a volume 4-form. If we assign

\begin{equation}
  \begin{array}{lr}
  \omega^0=\sqrt{1-\frac{2m}{r}}dt,\; \omega^1=\frac{1}{\sqrt{1-\frac{2m}{r}}},\\
  \omega^2=rd\theta,\;\omega^3=r\sin \theta d\phi,\\

  \end{array}
\end{equation}

we are going to deal with the dual forms of Schwarzschild spacetime. It is a good idea to learn math by examples. So here are some examples to find the dual forms. Ex1: Find $*(\omega^0\wedge \omega^1)$.

\begin{equation}
  \begin{array}{lr}
  $We may assume that $*(\omega^0\wedge \omega^1)=c(\omega^2\wedge \omega^3).\\
  $In 4 dimension spacetime, the dual of $dx\wedge dy$ must be $dt\wedge dz$ up to a constant$ \\
  $to be determined. $
  $But how to find c?$\\
  $We should put$
  (\omega^0\wedge \omega^1)\wedge*(\omega^0\wedge \omega^1)=|Matrix|*1\\
  $The martix is the inner product of dual form, that is,$
  \begin{pmatrix}
  \eta_{22} & 0  \\
  0 & \eta_{33}  \\

  \end{pmatrix}
  \\
  \rightarrow(\omega^0\wedge \omega^1)\wedge*(\omega^0\wedge \omega^1)
  =c(\omega^0\wedge \omega^1\wedge\omega^2\wedge \omega^3)\\
  =
   \begin{pmatrix}
  1 & 0  \\
  0 & 1  \\

  \end{pmatrix}*1.
  \therefore c=1.
  \end{array}
\end{equation}

We can try another examples, like $*(\omega^1\wedge \omega^2)=-1(\omega^0\wedge \omega^3)$. And furthermore $*(\omega^2)=c(\omega^0\wedge \omega^1\wedge \omega^3)$ by

\begin{equation}
  \begin{array}{lr}

  \omega^2\wedge*(\omega^2)=c\omega^2\wedge(\omega^0\wedge \omega^1\wedge \omega^3)\\
  =c\omega^0\wedge \omega^1\wedge \omega^2\wedge\omega^3)\\
  =
  \begin{pmatrix}
  -1 & 0 &0  \\
  0 & +1 &0  \\
  0 & 0 & +1 \\

  \end{pmatrix}*1=-*1\\

  \therefore c=-1.
  \end{array}
\end{equation}

In this way, we are able to construct

\begin{equation}
  \begin{array}{lr}
*(rd\theta)=-\sqrt{1-\frac{2m}{r}}dt\wedge\frac{1}{\sqrt{1-\frac{2m}{r}}}d\theta\wedge r\sin \theta d\phi\\
\rightarrow*(d\theta)=-\sin \theta dt\wedge dr\wedge d\phi, $ and another dual forms$.

  \end{array}
\end{equation}

And use ansatz like $F=F_1dt\wedge dr+F_2 d\theta \wedge d\phi$ to find $dF$ and $d*F$. This method is generally used in Einstein-Maxwell equations.

\appndxsec{A Summary of NP Formalism}
{\label{app:importantnotes}}

\begin{large}
\begin{equation}
    \begin{array}{lr}
\\
         dl=-(\epsilon+\bar{\epsilon})l\wedge n+(\alpha+\bar{\beta}-\bar{\tau})l\wedge m+(\bar{\alpha}+\beta-\tau)l\wedge\bar{m}\\
         -\bar{\kappa}n\wedge m-\kappa n\wedge\bar{m}
         +(\rho-\bar{\rho})m\wedge\bar{m}\\

         dn=-(\gamma+\bar{\gamma})l\wedge n+\nu l\wedge m+\bar{\nu} l\wedge\bar{m}-(\alpha+\bar{\beta}-\pi)n\wedge m\\
         -(\bar{\alpha}+\beta-\bar{\pi}) n\wedge\bar{m}
         +(\mu-\bar{\mu})m\wedge\bar{m}\\

         dm=-(\tau+\bar{\pi})l\wedge n+(\bar{\mu}+\gamma-\bar{\gamma})l\wedge m+\bar{\lambda} l\wedge\bar{m}\\
         -(\rho-\epsilon+\bar{\epsilon})n\wedge m-\sigma n\wedge\bar{m}
         +(\beta-\bar{\alpha})m\wedge\bar{m}\\
         \\
         \gamma_0=\gamma l+\epsilon n-\alpha m-\beta\bar{m}\\

         \gamma_1=-\tau l-\kappa n+\rho m+\sigma\bar{m}\\

         \gamma_2=\nu l+\pi n-\lambda m-\mu\bar{m}\\
         \\
         \theta_0=d\gamma_0-\gamma_1\wedge\gamma_2\\

         \theta_1=d\gamma_1-2\gamma_0\wedge\gamma_1\\

         \theta_2=d\gamma_2+2\gamma_0\wedge\gamma_2\\
         \\

         \\
    \end{array}
\end{equation}
\end{large}

\begin{large}
\begin{equation}
    \begin{array}{lr}

 \theta_0=(\Psi_2-\Lambda)(m\wedge\bar{m}-l\wedge n)+\Psi_3l\wedge m-\Psi_1n\wedge\bar{m}-\Phi_{11}(l\wedge n+m\wedge\bar{m})\\
         +\Phi_{12}l\wedge\bar{m}-\Phi_{10}n\wedge m\\

         \theta_1=\Psi_1(l\wedge n- m\wedge\bar{m})-(\Psi_2+2\Lambda)l\wedge m+\Psi_0n\wedge\bar{m}+\Phi_{01}(l\wedge n+m\wedge\bar{m})\\
         -\Phi_{02}l\wedge\bar{m}+\Phi_{00}n\wedge m\\

         \theta_2=\Psi_3(m\wedge\bar{m}-l\wedge n)+\Psi_4l\wedge m-(\Psi_2+2\Lambda )n\wedge\bar{m}-\Phi_{21}(l\wedge n+m\wedge\bar{m})\\
         +\Phi_{22}l\wedge\bar{m}-\Phi_{20}n\wedge m\\

    \end{array}
\end{equation}
\end{large}

\begin{large}
\begin{equation}
    \begin{array}{lr}

         4\pi T_{\mu\nu}=\Phi_{00}n_\mu n_\nu+\Phi_{22}l_\mu l_\nu+\Phi_{02}\bar{m}_\mu\bar{m}_\nu+\Phi_{20}m_\mu m_\nu\\
         -\Phi_{01}n_\mu\bar{m}_\nu-\Phi_{10}n_\mu m_\nu
         -\Phi_{12}l_\mu\bar{m}_\nu-\Phi_{21}l_\mu m_\nu\\
        +(\Phi_{11}+3\Lambda)(l_\mu n_\nu+n_\mu l_\nu)+(\Phi_{11}-3\Lambda)(m_\mu\bar{m}_\nu+\bar{m}_\mu m_\nu)\\

         \\
         G_{\mu\nu}=R_{\mu\nu}-\frac{R}{2}g_{\mu\nu}=-8\pi T_{\mu\nu}\\
         \\
         R=8\pi T=24\Lambda, 4\pi T=12\Lambda\\
         \\
         -\frac{1}{2}R_{\mu\nu}=\Phi_{00}n_\mu n_\nu+\Phi_{22}l_\mu l_\nu+\Phi_{02}\bar{m}_\mu\bar{m}_\nu+\Phi_{20}m_\mu m_\nu\\
         -\Phi_{01}n_\mu\bar{m}_\nu-\Phi_{10}n_\mu m_\nu
         -\Phi_{12}l_\mu\bar{m}_\nu-\Phi_{21}l_\mu m_\nu\\ 
         +(\Phi_{11}-3\Lambda)(l_\mu n_\nu+n_\mu l_\nu)+(\Phi_{11}+3\Lambda)(m_\mu\bar{m}_\nu+\bar{m}_\mu m_\nu)\\

    \end{array}
\end{equation}
\end{large}

\appndxsec{How to Derive the NPs , Riccis, and Weyls by an Example}{\label{app:importantnotes}}

\begin{equation*}
  \begin{array}{lr}
  \\
    ds^2=2dudv-(F(v)dx)^2-(G(v)dy)^2
    $Set$\;l=du,\;n=dv,\;\\
    m=\frac{1}{\sqrt{2}}(Fdx+iGdy),\;\bar{m}=\frac{1}{\sqrt{2}}(Fdx-iGdy)  \\
    dl=0=dn\\
    dx=\frac{m+\bar{m}}{\sqrt{2}F}\\
    dy=\frac{1}{i\sqrt{2}G}(m-\bar{m})\\
    $then$\\
    dm=\frac{1}{2}(\frac{F_\nu}{F}+\frac{G_\nu}{G})n\wedge m+\frac{1}{2}(\frac{F_\nu}{F}-\frac{G_\nu}{G})n\wedge \bar{m}.\\
    $See the summary of NP formalism, we have$\\
    -(\rho-\epsilon+\bar{\epsilon})=\frac{1}{2}(\frac{F_\nu}{F}+\frac{G_\nu}{G})\\
    -\sigma=\frac{1}{2}(\frac{F_\nu}{F}-\frac{G_\nu}{G})\\
    $The connection 1-form are$\\
    \gamma_1=\rho m+\sigma \bar{m}, \;\gamma_0=\gamma_2=0.\\
    $Such that,$\\
    d\gamma_1=\theta_1=-\frac{1}{2}n\wedge m(\frac{F_{\nu\nu}}{F}+\frac{G_{\nu\nu}}{G})-\frac{1}{2}n\wedge \bar{m}(\frac{F_{\nu\nu}}{F}-\frac{G_{\nu\nu}}{G})\\
    =\Phi_{00}n\wedge m+\Psi_0n\wedge \bar{m}.\\
    $Therefore the only non-zero Ricci and Weyl are $\\
    \Phi_{00}=-\frac{1}{2}(\frac{F_{\nu\nu}}{F}+\frac{G_{\nu\nu}}{G}),
    \Psi_0=-\frac{1}{2}(\frac{F_{\nu\nu}}{F}-\frac{G_{\nu\nu}}{G}).

  \end{array}
\end{equation*}

For pure EM wave, we may set $F=G$ to make $\Psi_0=0$, and let $F=cos(a\nu\theta(\nu))$. Then

\begin{equation}
  \begin{array}{lr}
    \Phi_{00}=-\frac{F_{\nu\nu}}{F} \\
    F_\nu=-a\theta(\nu)sin(a\nu\theta(\nu))\\
    F_{\nu\nu}=-a^2\theta(\nu)cos(a\nu\theta(\nu))\\
    \rightarrow\Phi_{00}=a^2\theta(\nu). $ We got the EM shock waves.$
  \end{array}
\end{equation}

\appndxsec{R, Weyls and Riccis}{\label{app:importantnotes}}

In the null-tetrad basis of NP, our tetrad components that are not zero are as follow

\begin{equation}
    \begin{array}{lr}
         \Lambda=\frac{R}{24}=\frac{kab\theta(u)\theta(v)}{3(1-2k)X^2}\\

         \Phi_{02}=\phi_{20}=-\frac{kab\theta(u)\theta(v)}{(1-2k)X^2}\\

         \Phi_{22}=\frac{a^2\delta(u)}{2X^2}[\tan(bv\theta(v))-\frac{1}{\sqrt{1-2k}}\tan(\frac{bv\theta(v)}{\sqrt{1-2k}})]+\frac{ka^2\theta(u)}{(1-2k)X^2}\\

         \Phi_{00}=\frac{b^2\delta(v)}{2X^2}[\tan(au\theta(u))-\frac{1}{\sqrt{1-2k}}\tan(\frac{au\theta(u)}{\sqrt{1-2k}})]+\frac{kb^2\theta(v)}{(1-2k)X^2}\\

         \Psi_2=\frac{ab\theta(u)\theta(v)}{3X^3}(\frac{kX}{1-2k}+3)\\

         \Psi_4=\frac{a\delta(u)}{2X^2}[\frac{1}{\cos(bv\theta(v))}+\frac{\cos(bv\theta(v))}{X}\\

         +\frac{1}{\sqrt{1-2k}}\tan(\frac{bv\theta(v)}{\sqrt{1-2k}})]-\frac{a^2\theta(u)}{X^3}(\frac{kX}{1-2k}+3)\\

         \Psi_0=\frac{b\delta(v)}{2X^2}[\frac{1}{\cos(au\theta(u))}+\frac{\cos(au\theta(u))}{X}\\

         +\frac{1}{\sqrt{1-2k}}\tan(\frac{au\theta(u)}{\sqrt{1-2k}})]-\frac{b^2\theta(v)}{X^3}(\frac{kX}{1-2k}+3)\\

    \end{array}
\end{equation}

where $X=1+\sin(au\theta(u)+bv\theta(v))$.

\end{appendices}

\printindex

\end{document}